\documentclass[12pt]{article}
\usepackage[all]{xy}
\usepackage{amssymb}
\makeatletter
\ifcase\@ptsize
 \font\teneufm=eufm10
 \font\seveneufm=eufm7
 \font\fiveeufm=eufm5
 \font\teneusm=eusm10
 \font\seveneusm=eusm7
 \font\fiveeusm=eusm5
\or
 \font\teneufm=eufm10 scaled \magstephalf
 \font\seveneufm=eufm7
 \font\fiveeufm=eufm5
 \font\teneusm=eusm10 scaled \magstephalf
 \font\seveneusm=eusm7
 \font\fiveeusm=eusm5

\or
 \font\teneufm=eufm10 scaled \magstep1
 \font\seveneufm=eufm7
 \font\fiveeufm=eufm5
 \font\teneusm=eusm10 scaled \magstep1
 \font\seveneusm=eusm7
 \font\fiveeusm=eusm5
\fi

\newfam\eufmfam
\newfam\eusmfam
\textfont\eufmfam=\teneufm  \scriptfont\eufmfam=\seveneufm
  \scriptscriptfont\eufmfam=\fiveeufm
\textfont\eusmfam=\teneusm  \scriptfont\eusmfam=\seveneusm
  \scriptscriptfont\eusmfam=\fiveeusm

\def\frak{\ifmmode\let\next\frak@\else
 \def\next{\errmessage{Use \string\frak\space only in math mode}}\fi\next}
\def\frak@#1{{\frak@@{#1}}}
\def\frak@@#1{\fam\eufmfam#1}

\def\sh{\ifmmode\let\next\sh@\else
 \def\next{\errmessage{Use \string\sh\space only in math mode}}\fi\next}
\def\sh@#1{{\sh@@{#1}}}
\def\sh@@#1{\fam\eusmfam#1}

\ifcase\@ptsize
 \font\tenmsa=msam10
 \font\sevenmsa=msam7
 \font\fivemsa=msam5
 \font\tenmsb=msbm10
 \font\sevenmsb=msbm7
 \font\fivemsb=msbm5
\or
 \font\tenmsa=msam10 scaled \magstephalf
 \font\sevenmsa=msam7
 \font\fivemsa=msam5
 \font\tenmsb=msbm10 scaled \magstephalf
 \font\sevenmsb=msbm7
 \font\fivemsb=msbm5
\or
 \font\tenmsa=msam10 scaled \magstep1
 \font\sevenmsa=msam7
 \font\fivemsa=msam5
 \font\tenmsb=msbm10 scaled \magstep1
 \font\sevenmsb=msbm7
 \font\fivemsb=msbm5
\fi

\newfam\msafam
\newfam\msbfam
\textfont\msafam=\tenmsa  \scriptfont\msafam=\sevenmsa
  \scriptscriptfont\msafam=\fivemsa
\textfont\msbfam=\tenmsb  \scriptfont\msbfam=\sevenmsb
  \scriptscriptfont\msbfam=\fivemsb

\def\Bbb{\ifmmode\let\next\Bbb@\else
 \def\next{\errmessage{Use \string\Bbb\space only in math mode}}\fi\next}
\def\Bbb@#1{{\Bbb@@{#1}}}
\def\Bbb@@#1{\fam\msbfam#1}
\def\hexnumber@#1{\ifnum#1<10 \number#1\else
 \ifnum#1=10 A\else\ifnum#1=11 B\else\ifnum#1=12 C\else
 \ifnum#1=13 D\else\ifnum#1=14 E\else\ifnum#1=15 F\fi\fi\fi\fi\fi\fi\fi}
\def\msa@{\hexnumber@\msafam}
\def\msb@{\hexnumber@\msbfam}
\mathchardef\square="0\msa@03

\makeatother
\newcommand{\beq}{\begin{equation}}
\newcommand{\eeq}{\end{equation}}
\newcommand{\ba}{\begin{array}}
\newcommand{\ea}{\end{array}}
\newcommand{\bea}{\begin{eqnarray}}
\newcommand{\eea}{\end{eqnarray}}
\newcommand{\bean}{\begin{eqnarray*}}
\newcommand{\eean}{\end{eqnarray*}}

\newtheorem{theorem}{Theorem}[section]
\newtheorem{prop}[theorem]{Proposition}
\newtheorem{lem}[theorem]{Lemma}
\newtheorem{defi}[theorem]{Definition}
\newtheorem{remark}[theorem]{Remark}

\newtheorem{proof}{Proof.}
\newenvironment{pf}{\begin{proof} \rm}{\hfill$\square$ \end{proof}}

\makeatletter
\@addtoreset{equation}{section}

\makeatother

\newcommand{\rref}[1]{(\ref{#1})} 

\def\v#1{{v^{(#1)}}}

\def\endpf{\begin{flushright}$\square$\end{flushright}}

\newcommand{\fraksl}{{\frak s}{\frak l}}

\def\res{\mbox res}

\def\sl23{{\fraksl}_3^{(2)}}
\begin{document}
\begin{titlepage}
\begin{center}
{\huge
New Integrable Hierarchies from  Vertex Operator Representations of Polynomial
Lie Algebras}
\end{center}
\vspace{0.8truecm}
\begin{center}
{\large
Paolo Casati$^\diamondsuit$, Giovanni Ortenzi$^\sharp$
\vskip0.8truecm
Dipartimento di Matematica e applicazioni\\
II Universit\`a di Milano\\
Piazza dell'Ateneo Nuovo 1, I-20126 Milano, Italy\\
E--mail$^\diamondsuit$:  casati@matapp.unimib.it\\
E--mail$^\sharp$:  giolev@matapp.unimib.it},
\end{center}
\vspace{0.2truecm}
\vspace{0.2truecm}
\abstract{\noindent
We give a representation--theoretic interpretation 
of recent discovered coupled soliton equations
using   vertex operators  construction of affinization of not simple 
but quadratic Lie algebras. In this setup we are able to obtain 
new integrable hierarchies coupled to each Drinfeld--Sokolov of $A$, $B$, 
$C$, $D$ hierarchies and to construct their soliton solutions.}\vskip 2truecm \noindent
\end{titlepage}

\section{Introduction}

\bigskip
One of the most important achievement of the 
representations of the affine Lie algebras  and their groups is surely the 
Lie theoretical explanation of the Hirota \cite{H} bilinear approach to the 
soliton equations. 
This beautiful piece of mathematics is the result of an important sequence 
of relevant  papers, which starts in 1981 with the works of Sato \cite{S}  \cite{SS},
where the link between the soliton equations and the infinite--dimensional groups  has been 
brought to the light for the first time. Some years  later Date, Jimbo, Kashiwara and Miwa
gave a construction of the Kadomtsev--Petviashvili (KP) and Korteweg--de Vries
(KdV) hierarchies in terms of the vertex operators representating 
the affine lie algebra ${\frak a}_\infty$ and $\widehat{{\frak s}{\frak l}}_2$ 
respectively, while Segal and Wilson \cite{SW} have examinated the same equation from 
a geometrical point of view. Finally Drinfeld, Sokolov \cite{DS},  Kac Peterson  
and Wakimoto \cite{K} \cite{KW}\cite{KP}  
have extended this theory to all  affine Lie algebras.
These results have suggested to find a similar interpretation for  others 
hierarchies of soliton equations, for example in a recent work \cite{B} Billig 
has obtained this goal for the sine--Gordon.\par
The main aim of the present paper is to contribute to the research in this direction.
Our starting point is the following ``coupled KdV equations'' 
which appears in many  very recent papers of different authors like 
 Hirota,  Hu,  Tang,  \cite{HXT}, Sakovich \cite{Sak},  Kakei \cite{SK}:
\beq
\label{iKdV2}
\begin{array}{ll}
v_{t}+6vv_{x}+v_{xxx}=0&\\
w_t+6vw_x+w_{xxx}=0.&
\end{array}
\eeq
The corresponding bilinear Hirota form  of these equations (and actually of many others
closely related, among them  ``coupled KP equations'' \cite{Sak}) is namely known \cite{HXT}
together with some soliton solutions, but, as far as we know, it is still missing
their broader Lie--theoretic interpretation.  In this paper we shall show how these equations
are a particular case of a very wide class of ``coupled soliton equations'' 
which can be obtained using the vertex operator realization of a new class 
of infinite dimensional Lie algebras. 
These latter algebras are the affinization of not simple finite dimensional Lie algebras,
which still posses a symmetric non degenerated ad--invariant bilinear form.
Therefore in our long journey towards our task we shall be
enforced to develop a vertex operator algebras theory for a class of Lie algebras
which are not the the affinization of semisimple ones. 
Nevertheless our construction will allow us to produce 
coupled soliton equations corresponding to each of the Drinfeld--Sokolov and the 
AKP BKP CKP DKP hierarchies, although for sake of brevity only the case of the 
coupled AKP BKP and their reductions to opportune generalisations of the 
affine Lie algebras $A_1^{(1)}$ $A_2^{(1)}$,  $A_1^{(2)}$ and 
and  $B_2^{(1)}$ are explicitly examinated in the paper. These reductions in turn
provide a clear explanation of how the coupled KP equations   become the coupled KdV 
ones by neglecting the dependency from one particular variable.
Finally  the action on the space of representation
of the corresponding infinite dimensional groups will provide, exactly as in the usual 
case, a class of multi--soliton solutions.\par  
 The paper is organised as follows: in the second section we shall describe a
class of finite dimensional Lie algebras known in the literature as
polynomial Lie algebras
\cite{MM},\cite{PV} which, roughly speaking, can be regarded as direct sum of semisimple Lie 
algebras
endowed with a non canonical Lie bracket. We shall show that these Lie algebras
can be constructed in  completely different ways: namely as particular finite
dimensional quotients of an infinite dimensional algebra and as a
Wigner contraction  of a direct sum of finite
dimensional semisimple Lie algebras,  or finally as tensor product between a finite dimensional
 Lie algebra $\frak g$
and a nilpotent commutative ring. Further we shall show how on these Lie algebras is defined a
class of symmetric non degenerated ad--invariant bilinear forms if a such bilinear form
exits on $\frak g$.
In the next section it will be shown how these latter bilinear forms can be used to affinize 
those in general non simple Lie algebras. Then in section 4 their vertex
operator algebras construction is presented. Once this result is achieved we
can tackle the problem to construct the corresponding generalised Hirota
bilinear equation and their multisoliton solution in term of
$\tau$--functions. This will be done in the fifth and last section 
where further the case of the coupled AKP BKP and their reduction to Lie algebras 
generalising the algebras $A^{(1)}_1$, $A^{(1)}_2$, $A^{(2)}_1$, and $B^{(1)}_2$ 
are presented into details.

\section{The polynomial Lie algebras}

The aim of this first section is to construct in a few different ways
a class of finite dimensional Lie algebras (called in what follows polynomial
Lie algebras to keep the name usually used in the literature, see for example
\cite{MM} and \cite{PV}) which are going to play a crucial role in the whole 
paper.
\noindent The reader will surely recognise  that actually our constructions can
be  
straightforward generalised to the realm of the infinite dimensional Lie
algebras; but, although we shall need for our porpoises this generalisation,
let us here for sake of concreteness restrict ourself to the maybe simpler finite
dimensional case.

Let $\frak g$ be a finite dimensional Lie algebra over $\Bbb C$ and let us
denote by $\Bbb C (\lambda)$ the ring of the polynomials in a complex variable
$\lambda$ with coefficients in $\Bbb C$:
\beq
\Bbb C (\lambda)=\{\sum_{i=0}^kc_{i}\lambda^{i}\vert\ k \in\Bbb N\}.
\eeq

The space $\Bbb C (\lambda)$ of course inherits form its associative and
commutative product an abelian (i.e. trivial) Lie algebra structure. Therefore
we can regard the space
\beq
\frak g(\lambda)=\frak g \otimes\Bbb C (\lambda)
\eeq
as a Lie algebra tensor product. It may be identified with the Lie algebra
of polynomial maps ${\Bbb C} \to \frak g$, hence an element $X(\lambda)$ in $\frak g(\lambda)$
can be viewed as the mapping $X:\Bbb C\to \frak g$,
$X(\lambda)=\sum_{k=0}^{\infty}X_{k}{\lambda}^{k}$ where $X_{k}=0$ for all but
a finite number of indices. In this setting the Lie bracket of two elements in   
${\frak g} (\lambda)$, $X(\lambda)=\sum_{k=0}^{\infty}X_{k}{\lambda}^{k}$ and 
$Y(\lambda)=\sum_{k=0}^{\infty}Y_{k}{\lambda}^{k}$ can be written explicitly as
\beq
[X(\lambda),Y(\lambda)]=\sum_{n=0}^{\infty}(\sum_{j=0}^{n}[X_{j},Y_{n-j}%
]_{\frak g})\lambda^{n}%
\eeq
where $[\cdot,\cdot]_{\frak g}$ is the Lie bracket defined on $\frak g$.\par
If on $\frak g$ is defined an ad--invariant bilinear form $\langle\cdot
,\cdot\rangle$ (for instance the Killing form) then we can carry it to a
bilinear form $\langle\cdot,\cdot\rangle_{\lambda}$ on $\frak{g}(\lambda)$ by setting on $\Bbb
C (\lambda)$ the standard inner product
\beq
\label{bfmm}
(p(\lambda),q(\lambda))=(\sum_{i=0}^{\infty}p_{i}\lambda^{i},\sum
_{j=0}^{\infty}q_{j}\lambda^{j})= \int_{\vert\lambda\vert=1} \lambda
^{-1}p(\lambda)\overline{q(\lambda)}d\lambda=\sum_{i=0}^{\infty}p_{i}\overline{q_{i}}%
\eeq
(where $\overline{z}$ denotes the complex conjugate of $z$ and 
one has to keep in mind that all the sum are on a finite set and that
on the unit circle both $p(\lambda)$ and $p(\lambda^{-1})$ are well defined).
More precisely on $\frak g(\lambda)$ will be defined the bilinear form:
\bea
\label{inprogl}
\langle X(\lambda),Y(\lambda)\rangle_{\lambda}&=& \langle \sum_{i=0}^{\infty}
X_{i}\lambda^{i},\sum_{j=0}^{\infty}Y_{j}\lambda^{j} \rangle_{\lambda}\\
 &=& \int_{\vert\lambda
\vert=1} \lambda^{-1} \langle X(\lambda),Y(\lambda^{-1}) \rangle d\lambda=\sum_{i=0}^{\infty
} \langle X_{i},Y_{i} \rangle. \nonumber
\eea
Unfortunately this bilinear form, while it is not degenerate if
the bilinear form on $\frak g$ is not degenerate, turns out to be in general
not ad--invariant even if the form chosen on $\frak g$ (like in the case of
the Killing form) is ad--invariant. Suppose indeed  $\frak g=\frak s\frak
l (2,\Bbb C)$ endowed with the Killing form and set
$$X=\left(\begin{array}{cc}
0 & 1\\
0 & 0
\end{array}
\right)  \quad H=\left(
\begin{array}{cc}
1 & 0\\
0 & -1
\end{array}
\right)  
\quad Y=\left(
\begin{array}{cc}
0 & 0\\
1 & 0
\end{array}
\right)
$$
then
$$
\langle H,[X\lambda,Y\lambda]\rangle_{\lambda}= \langle H,H\lambda^{2}
\rangle_{\lambda}=0
$$
but
$$
\langle[H,X\lambda],Y\lambda\rangle_{\lambda}= \langle2X\lambda,Y\lambda
\rangle_{\lambda}=2\neq0.
$$
On the other hand there exist on $\frak{g}(\lambda)$  non trivial  
 symmetric ad--invariant form, which are  surely more exotic than that 
previously considered and moreover degenerated.
Nevertheless these latter  bilinear forms are  still worth  to be considered, 
because from them  one can derive symmetric  ad--invariant bilinear forms 
on opportune quotients
of $\frak{g}(\lambda)$ which will turn out to be actually  non degenerated.
\begin{defi}
\label{glbilin}
For any symmetric bilinear form  $\langle \cdot, \cdot \rangle_{\frak g}$  on $\frak{g}$ and for 
any  sequence ${\cal A}=\{ a_n \}_{n\in \Bbb N}$  of complex numbers $a_k$, 
let $\langle \cdot, \cdot \rangle_{\cal A}$ be the bilinear form on $\frak{g}(\lambda)$ given by the formula:
\bea
\label{glbilinform}
\langle \cdot, \cdot \rangle_{\cal A}:
\frak{g}(\lambda) \times \frak{g}(\lambda) &\longrightarrow& {\Bbb C} \nonumber \\
X(\lambda),Y(\lambda) \in \frak{g}(\lambda) &\longrightarrow& \langle X(\lambda),Y(\lambda)  \rangle_{\cal A} =
\sum_{n=0}^{\infty}a_n \sum_{k=0}^{n} \langle X_{n-k},Y_{k} \rangle_{\frak g}
\eea
\end{defi}
It easily checked the following:
\begin{prop}
\label{glbilinprop}
If the symmetric bilinear form $\langle \cdot, \cdot \rangle_{\frak g}$ defined on $\frak{g}$ is 
ad--invariant then  
the same is also true for the bilinear form \rref{glbilinform} defined on 
$\frak{g}(\lambda)$.
\end{prop}
{\bf Proof.}
We have to prove that for any elements 
$X(\lambda)=\sum_{n=0}^{\infty} X_n \lambda^n$, $Y(\lambda)=\sum_{k=0}^{\infty} Y_j \lambda^j$,
 $Z(\lambda)=\sum_{k=0}^{\infty} Z_k \lambda^k$ in $\frak{g}(\lambda)$ it holds that 
\beq
\label{adgl}
\langle [X(\lambda),Y(\lambda)] , Z(\lambda) \rangle_{\cal A} 
= \langle X(\lambda),[Y(\lambda) , Z(\lambda)] \rangle_{\cal A}. 
\eeq
Now
\bea
\label{adgl1}
\langle [X(\lambda),Y(\lambda)] , Z(\lambda) \rangle_{\cal A} &=& 
\sum_{n=0}^{\infty} a_n \sum_{k=0}^{n} \langle [X,Y]_{n-k},Z_k \rangle_{\frak g} \\ &=&
\sum_{n=0}^{\infty} a_n \sum_{k=0}^{n} \langle \sum_{j=0}^{n-k} [X_j,Y_{n-k-j}],Z_k \rangle_{\frak g} 
\nonumber
\eea
while
\bea
\label{adgl2}
\langle X(\lambda),[Y(\lambda) , Z(\lambda)] \rangle_{\cal A} &=&
\sum_{n=0}^{\infty} a_n \sum_{k=0}^{n} \langle X_{n-k},[Y,Z]_k \rangle_{\frak g}  \\ &=&
\sum_{n=0}^{\infty} a_n \sum_{k=0}^{n} \langle X_{n-k},\sum_{j=0}^{k} [Y_k,Z_{k-j}] \rangle_{\frak g}. 
\nonumber
\eea
To see that \rref{adgl} holds if suffices to observe that both 
\rref{adgl1} and \rref{adgl2}
can be written as
\bean
\langle [X(\lambda),Y(\lambda)] , Z(\lambda) \rangle_{\cal A} &=&  
\sum_{n=0}^{\infty}a_n \sum_{j_1+j_2+j_3=n}\langle [X_{j_1},Y_{j_2}],Z_{j_3} \rangle_{\frak g} \\
\langle X(\lambda),[Y(\lambda) , Z(\lambda)] \rangle_{\cal A} &=& 
\sum_{n=0}^{\infty} a_n \sum_{j_1+j_2+j_3=n}\langle X_{j_1},[Y_{j_2},Z_{j_3}] \rangle_{\frak g}
\eean
and that the equation \rref{adgl} immediately follows using the 
ad--invariance of the bilinear form $\langle\cdot,\cdot \rangle_{\frak g}$.
\endpf
As already remarked the bilinear form \rref{glbilinform} is   degenerated
 for every sequence ${\cal A}=\{a_n\}_{n\in \Bbb N}$. It is namely 
easily checked that its radical is given by the subspace of particular  
``polynomials'' in $\frak g(\lambda)$   which vanish  in $\lambda=1$. 
This is due to the fact that our bilinear map
can be factorized in a bilinear map from 
$\frak{g}(\lambda) \times \frak{g}(\lambda)$ to ${\Bbb C}(\lambda)$ 
and an ``evaluation''  map (in $\lambda=1$) from ${\Bbb C}(\lambda)$ to  ${\Bbb C}$.
It is matter of fact that these two maps commute in an appropriate 
but straightforward sense,
and while the first map is not degenerate the second one has of course always 
a non trivial kernel. For example, in the most interesting case, when $a_n=1$ for all $n$,
the kernel coincides with the linear space of all polynomials $X(\lambda)$ 
which vanish in $\lambda=1$.\\
For what we have in mind the most important property of the infinite
dimensional Lie algebra $\frak g (\lambda)$ is that it has an infinite
number of nested ideals. Namely for each $k \in\Bbb N$ let us consider the
subspace $I_{k}(\lambda)$ of all elements of $\frak g(\lambda)$ which can be
written as the product $\lambda^{k+1}Y(\lambda)$ for some $Y(\lambda)$ in $\frak g(\lambda)$ i.e.,
\beq
\label{lieideal}
I_{k}(\lambda)=\{X(\lambda)\in g(\lambda)\vert\ X(\lambda)=\lambda^{k+1}
Y(\lambda) \quad Y(\lambda)\in g(\lambda)\}.
\eeq
From (\ref{lieideal}) it follows immediately that $I_{k}(\lambda)$ is for every
$k$ an ideal and that $I_{h}(\lambda)\subset I_{k}(\lambda)$ if and only if
$h\geq k$. Using this property we can define our polynomials Lie algebras:
\begin{defi}
\label{pla1d} The polynomial Lie algebra over $\frak g$ of degree 
$n$ $\frak g^{(n)}(\lambda)$ is the quotient Lie algebra
\beq
\frak g^{(n)}(\lambda)=\frak g(\lambda)/I_n(\lambda).
\eeq
\end{defi}
Since the canonical projection $\pi_{n}:\frak g(\lambda)\to\frak g
^{(n)}(\lambda)$ acts on a generic element $Z(\lambda)$ of $\frak g(\lambda)$
as
\beq
\label{canproj}
\pi_{n}(Z(\lambda))=\pi_{n}(\sum_{j=0}^{\infty}Z_{j}\lambda^{j})=\sum_{j=0}^{n}
Z_{j}\lambda^{j}
\eeq
we can identify $\frak g^{(n)}(\lambda)$ with the vector space of all
polynomials in $\lambda$ with coefficients in $\frak g$ of degree less or
equal to $n$ endowed with the Lie bracket:
\beq
\label{liepro1}
\left[  \sum_{j=0}^{n}X_{j}\lambda^{j},\sum_{i=0}^{n}Y_{i}\lambda^{i}\right]
= \sum_{k=0}^{n} \sum_{j=0}^{k} [X_{j},Y_{k-j}]\lambda^{k}.
\eeq
This latter formula suggests an alternative and more direct but maybe less
natural definition of $\frak g^{(n)}(\lambda)$:
\begin{defi}
\label{IIdgn} Let $\frak g$ any finite dimensional Lie algebra, then we will
denote by $\frak g^{(n)}$ the Lie algebra isomorph as vector space to the
direct sum ${\frak g}^n=\oplus_{i=0}^{n} \frak g_{i}$, $\frak g_{i}\simeq\frak g$ of $n+1$
copies of $\frak g$ and with Lie bracket given by the
relation
\beq
\label{lbmm}
\begin{array}{ll}
 [(X_{0},\dots,X_{n}),(Y_{0},\dots,Y_{n})]= & (Z_{0},\dots
,Z_{n})\\
Z_{k}=\sum_{j=0}^{k}[X_{j},Y_{k-j}] & k=0,\dots,n.
\end{array}
\eeq
\end{defi}
We should point out here that the Lie algebras $\frak g^{(n)}$, which is obviously 
equivalent to ${\frak g}^{(n)}(\lambda)$,  was already
defined and considered in \cite{MM} (in particular definition \ref{IIdgn}
coincides with the definition of the polynomial Lie algebras with bracket
$[\cdot,\cdot]_{0}$ (in their notation) contained in their proposition 4.4).
From this definition one would natural define on $\frak g^{(n)}$ the symmetric 
bilinear form  
\beq
\label{bfmm1}
\langle(X_{0},\dots,X_{n}),(Y_{0},\dots,Y_{n})\rangle=\sum_{i=0}^{n}\langle
X_{i}, Y_{i} \rangle
\eeq
i.e., the bilinear form canonically defined on the direct sum 
$\oplus_{i=0}^{n} \frak g_{i}$, which is also the projection over $\frak g^{(n)}$ 
of the symmetric bilinear form  \rref{bfmm} defined on $\frak g(\lambda)$ and which is 
usually considered in the literature \cite{MM}, \cite{FMPZ}.
But clearly being this symmetric bilinear form ad--invariant for the  brackets
of  $\oplus_{i=0}^{n} \frak g_{i}$, it fails to be ad--invariant for the modified ones
\rref{lbmm}. 
Therefore one may wonder 
if on $\frak{g}^{(n)}$ there exist symmetric non degenerated bilinear 
ad--invariant forms. Since for $n>0$ the Lie algebra 
$\frak{g}^{(n)}$ is not semisimple or reductive, even   if the Lie algebra
$\frak{g}$ has  this property, as a consequence of the Cartan theorem \cite{Hu} 
we cannot use the usual Killing form, and we must look for something else.
 As we shall show in a moment the bilinear forms, we are looking for, will 
be deduced  from the symmetric ad--invariant bilinear forms defined in 
\ref{glbilin}.  
Unfortunately to ``project''  these latters  on the quotient space
$\frak{g}^{(n)}$ is a little more cumbersome because the inner product of two 
representatives of two equivalence classes of the equivalence relation induced 
by $I_{n}(\lambda)$ depends in general from the representatives themselves and not 
only from their classes. Nevertheless it is possible to define on 
$\frak{g}^{(n)}$ bilinear form which can be regarded as ``natural projection'' 
on $\frak{g}^{(n)}$ of the bilinear forms \rref{glbilinform} defined on 
$\frak{g}(\lambda)$. Of course these
forms on $\frak{g}^{(n)}$ can be also defined independently without any referring 
to \rref{glbilinform}.\\
Although it should be easy for the reader to guess how these bilinear forms
look like, let us construct them carefully to show into details their relation 
with \rref{glbilinform}.\\
As already pointed out the bilinear form \rref{glbilinform} \\ \noindent 
$\Omega_{\cal A}(X(\lambda),Y(\lambda))=\langle(X(\lambda),Y(\lambda)\rangle_{\cal A} $ 
can be factorized as follows:\\
\beq
\xymatrix{{\frak{g}(\lambda) \times \frak{g}(\lambda)} \ar[rr]^-{\Omega^{\lambda}_{\cal A}} 
\ar@/_2pc/[rrrr]_{\Omega_{\cal A}}
&& {\Bbb C}(\lambda) \ar[rr]^{ev} && {\Bbb C}\\
 }
\eeq
where $\Omega^{\lambda}_{\cal A}:\frak{g}(\lambda) \times \frak{g}(\lambda) \longrightarrow {\Bbb C }$ 
is the map:
\beq
\label{omegalambda}
X(\lambda),Y(\lambda) \in \frak{g}(\lambda) \longrightarrow \Omega^{\lambda}_{\cal A} ( X(\lambda),Y(\lambda)) =
\sum_{n=0}^{\infty} a_n \sum_{k=0}^{n} \langle X_{k},Y_{n-k} \rangle_{\frak g} \lambda^n
\eeq
while $ ev:{\Bbb C}(\lambda) \longrightarrow {\Bbb C}$  is simply the map which evaluate
a polynomial at the point $\lambda=1$:
\beq
\label{2eval}
p(\lambda) \in {\Bbb C}^{(n)}(\lambda)\mapsto ev(p(\lambda))=p(1).
\eeq 
Now since obviously the subspace 
$${\cal I}_{n}(\lambda)=\{p(\lambda)\in{\Bbb C}
|\ \exists\ q(\lambda) \in {\Bbb C} : p(\lambda) = \lambda^{n+1} q(\lambda)  \} $$
is an ideal in the commutative algebra ${\Bbb C}(\lambda)$ we can consider the quotient 
space
\beq
\label{quotspc}
{\Bbb C}^{(n)}(\lambda): {\Bbb C}(\lambda)/{\cal I}_{n}(\lambda)
\eeq
and the corresponding quotient map $\Pi^{n}$, which can be explicitly written as
\bea 
\label{pkclck}
\Pi^{n}:{\Bbb C}(\lambda) &\longrightarrow&{\Bbb C}^{(n)}(\lambda) \nonumber \\
\Pi^{n}\left(\sum_{j=0}^{\infty} p_j \lambda^j\right) &=&\sum_{j=0}^{n} p_j \lambda^j.
\eea
Then it is easily to show that it holds:
\begin{prop}
\label{omkl}
There exists a unique map 
$$\Omega^{(n)}_{\cal A}(\lambda):\frak{g}(\lambda) \times \frak{g}(\lambda) \longrightarrow {\Bbb C}^{(n)}(\lambda)$$
which makes the diagram : \\
\beq
\label{dglglk} 
\xymatrix{{\frak{g}(\lambda) \times \frak{g}(\lambda)}
\ar[rr]^{\Omega^{\lambda}_{\cal A}}\ar[dd]_{\pi_{n} \times \pi_{n} } && 
{{\Bbb C}(\lambda)} \ar[dd]^{\Pi^n} \\ 
&{\circlearrowleft}& \\
{\frak{g}^{(k)}(\lambda) \times \frak{g}^{(k)}(\lambda)} \ar[rr]_-{\Omega^{(n)}_{\cal A}(\lambda)} 
&& {{\Bbb C}^{(k)}(\lambda)}}\\
\eeq
commutative, where the maps $\Omega^{\lambda}_{\cal A}$, $\pi_{n}$ and $\Pi^n$ are respectively
 given in \rref{omegalambda},\rref{canproj} and \rref{pkclck}.
\end{prop}
{\bf Proof}
It is almost evident that the map
\bea
\Omega^{(n)}_{\cal A}(\lambda):\frak{g}^{(n)}(\lambda) \times \frak{g}^{(n)}(\lambda) &\longrightarrow& 
{\Bbb C }^{(n)}(\lambda) \nonumber \\
X(\lambda),Y(\lambda) \in \frak{g}^{(n)}(\lambda) &\mapsto& 
\Omega^{(n)}_{\cal A}(\lambda)( X(\lambda),Y(\lambda))  \\
& &= \sum_{j=0}^n a_j \sum_{i=0}^j \langle X_{i},Y_{j-i} \rangle_{\frak g} \lambda^j \nonumber
\eea
makes the diagram \rref{dglglk} commutative. To show the unicity it is enough
to check that if $X(\lambda)$ and $Y(\lambda)$ are two elements in $\frak{g}^{(n)}(\lambda)$
then for every  $\widetilde{X}(\lambda)$ and  $\widetilde{Y}(\lambda)$ respectively in 
${\pi_n}^{-1}(X(\lambda))$ and ${\pi_n}^{-1}(Y(\lambda))$ it holds:
\beq
\Omega_{\cal A}^{(n)}(\lambda)(X(\lambda),Y(\lambda))
=\Pi^n(\Omega^{\lambda}_{\cal A}(\widetilde{X}(\lambda),\widetilde{Y}(\lambda))).
\eeq
But this is a straightforward computation.
 Indeed any $\widetilde{X}(\lambda)$ and  $\widetilde{Y}(\lambda)$ will have the form:
\bea
\widetilde{X}(\lambda)=X(\lambda)+\sum_{j \geq n+1}c_j \lambda^j \nonumber \\
\widetilde{Y}(\lambda)=Y(\lambda)+\sum_{j \geq n+1}d_j \lambda^j \nonumber 
\eea
for some elements $c_j,d_j \in \frak{g}$, therefore
\bea
\Omega^{\lambda}_{\cal A}(\widetilde{X}(\lambda),\widetilde{Y}(\lambda)) &=&
\Omega^{\lambda}_{\cal A}(X(\lambda),Y(\lambda))+\Omega^{\lambda}_{\cal A}(X(\lambda),d(\lambda)) 
\nonumber \\
& & +\Omega^{\lambda}_{\cal A}(c(\lambda),Y(\lambda))+\Omega^{\lambda}_{\cal A}
(c(\lambda),d(\lambda)), \nonumber
\eea
with
\bea
\Omega^{\lambda}_{\cal A}(X(\lambda),Y(\lambda))
&=&\sum_{j=0}^{n}a_j\sum_{i=0}^{j}\langle X_i,Y_{j-i}\rangle_{\frak g} \lambda^j \nonumber \\
\Omega^{\lambda}_{\cal A}(X(\lambda),d(\lambda))
&=&\sum_{j=n+1}^{\infty}a_j\sum_{i=0}^{j}\langle X_i,d_{j-i}\rangle_{\frak g} \lambda^j \nonumber \\
\Omega^{\lambda}_{\cal A}(c(\lambda),Y(\lambda))
&=&\sum_{j=n+1}^{\infty}a_j\sum_{i=0}^{j}\langle c_{j-i},Y_{i}\rangle_{\frak g} \lambda^j \nonumber \\
\Omega^{\lambda}_{\cal A}(c(\lambda),d(\lambda))
&=&\sum_{j=2n+1}^{\infty}a_j\sum_{i=0}^{j}\langle c_i,d_{j-i}\rangle_{\frak g} \lambda^j. \nonumber 
\eea
Thus
\beq
\Pi^n(\Omega_{\cal A}^{\lambda}(\widetilde{X}(\lambda),\widetilde{Y}(\lambda)))=
\sum_{j=0}^{n}a_j\sum_{i=0}^{j}\langle X_i,Y_{j-i}\rangle_{\frak g} \lambda^j=
\Omega^{(n)}_{\cal A}(\lambda)(X(\lambda),Y(\lambda)).
\eeq
\endpf
\begin{defi}
\label{glkbil}
Let us denote 
\beq
\Omega^{(n)}_{\cal A}:\frak{g}^{(n)}(\lambda) \times \frak{g}^{(n)}(\lambda) \longrightarrow {\Bbb C}
\nonumber
\eeq
the composition $ev^{(n)} \circ \Omega^{(n)}_{\cal A}(\lambda)$  where $\Omega^{(n)}_{\cal A}(\lambda)$ 
is the map defined in proposition \ref{omkl} and $ev^{(n)}$ is the ``evaluation'' in $\lambda=1$ map
on nilpotent commutative space ${\Bbb C}^{(n)}(\lambda)$ given by:
\bean
{\Bbb C}^{(n)}(\lambda) &\longrightarrow& {\Bbb C}  \\
p(\lambda) &\mapsto& ev^{(n)}(p(\lambda))=p(1). 
\eean 
In formula
\beq
\label{omlkbf}
\Omega^{(n)}_{\cal A}(X(\lambda),Y(\lambda))=\langle X(\lambda),Y(\lambda)\rangle^{(n)}_{\cal A}= \sum_{j=0}^{n}a_j
\sum_{i=0}^{j}\langle X_i,Y_{j-i}\rangle_{\frak g}.
\eeq
\end{defi}
Let us here  stress once again that one can use definition \ref{glkbil} 
and the formula \rref{omlkbf} without referring to the previous proposition 
or the bilinear form $\Omega^\lambda_{\cal A}$ and that obviously this bilinear form 
is also defined on the equivalent Lie algebra ${\frak g}^{(n)}$.\\
The maps $\Omega^{(n)}_{\cal A}$ are the bilinear forms which we was looking for,
\begin{prop}
\label{pseudokill}
For any $n \in {\Bbb N}$ and any set of complex numbers  
${\cal A}= \{a_j\}_{j=0,\dots n}$, the map 
$\Omega^{(n)}_{\cal A}:\frak{g}(\lambda) \times \frak{g}(\lambda) \longrightarrow {\Bbb C}$
is a bilinear, ad--invariant and, if $a_n\neq 0$, not degenerate form.
\end{prop} 
{\bf Proof }
That $\Omega^{(n)}_{\cal A}(\lambda)$ is a bilinear form follows immediately by its definition.\\
Further  to prove its ad--invariance one has only to repeat 
{\it mutata mutandis} exactly 
the proof given to show the same property for the bilinear form $\Omega^{\lambda}_{\cal A}$ in 
proposition \ref{glbilinprop}.\par\noindent 
It remains to show that it is non-degenerated. This can be done in two different ways.
First let us show it by proving that if $X(\lambda) \in \frak{g}^{(n)}$ is
such that 
$\langle X(\lambda), Y(\lambda)\rangle^{(n)}_{\cal A}=0 $  for every $ Y(\lambda) \in \frak{g}^{(n)} $  
then $X(\lambda)=0$. Indeed since $X(\lambda)$ has the form $X(\lambda)=\sum_{j=0}^{n}X_j \lambda^j$
its inner product with an element of the form $Y_n(\lambda)=Y_n \lambda^n$
will be:
\beq
\langle X(\lambda),Y_n(\lambda)\rangle^{(n)}_{\cal A}=a_n\langle X_0 , Y_n \rangle_{\frak g}
\nonumber
\eeq
and since $Y_n$ can be chosen arbitrarily in $\frak{g}$,  $\langle \cdot , \cdot \rangle_{\frak g}$ is 
non degenerated on it and $a_n\neq 0$,   $\langle X(\lambda),Y_n(\lambda)\rangle^{(n)}_{\cal A}=0$   
implies that $X_0=0$. Then by pairing $X(\lambda)$ with an element of the type 
$Y_{n-1}(\lambda)=Y_{n-1} \lambda^{n-1}$ we get:
\beq
\langle X(\lambda),Y_{n-1}(\lambda)\rangle^{(n)}_{\cal A}=a_n \langle X_1 , Y_{n-1} \rangle_{\frak g}
\nonumber
\eeq
and then again $\langle X(\lambda),Y_{n-1}(\lambda)\rangle^{(n)}_{\cal A}=0$  implies $X_1=0$. 
Repeating n times this argument we obtain step by step that each coefficient  
$X_i$ is zero, proving the proposition.\\
Actually passing through the matricial representation of $\Omega^{(n)}_{\cal A}$
we can give a more compact proof of this fact. 
If we denote by $\omega$ the matricial representation of the bilinear form 
$\langle \cdot , \cdot \rangle_{\frak g}$ defined on $\frak{g}$ then it is immediately to
show that $\Omega^{(k)}$ has the matricial form

\beq
\label{matrpseudokill}
\Omega^{(n)}_{\cal A}= \left(\begin{array}{c c c c c c}
                     a_0\omega &a_1\omega & a_2\omega & \dots & a_{n-1}\omega &a_n\omega \\
                     a_1 \omega & a_2\omega & & \dots & a_n\omega & 0 \\
                     a_2\omega &  &  & \dots & 0 & 0 \\
                     \vdots  &    & \;& \;& \;&\vdots \\
                     a_{n-1}\omega  & a_n\omega &0 & \dots  & 0 & 0 \\
                     a_n\omega &0 & 0& \dots & 0 & 0   
                    \end{array} \right).
\eeq
But then a simple argument by induction shows that 
\beq
{\rm det } (\Omega^{(n)}_{\cal A})={a_n}^n\det(\omega)^n  
\nonumber
\eeq
and therefore that $\Omega^{(n)}_{\cal A}$ is non degenerated if and only if
 $\omega$ is not degenerated. 
\endpf
The just proved proposition shows that the polynomial Lie algebra ${\frak g}^{(n)}$
when $\frak g$ is semisimple, are non trivial (non abelian or semisimple)
examples of quadratic Lie algebras
i.e., finite dimensional Lie algebras which possess a symmetric ad--invariant,
non degenerated bilinear form \cite{MER} and \cite{OO}. 
This fact may justify the restriction  in this section of 
our construction to the finite dimensional case.\par 
Let us rest here for a while to reconsider Definition \ref{IIdgn}.
This definition  may be reformulated by saying that the Lie algebra
$\frak g^{(n)}$ can be viewed as the direct sum $\frak g^n=\bigoplus_{k=0}^n{\frak g}_k$
of $n+1$ copies of $\frak g$ endowed with a maybe cumbersome Lie brackets.
It is therefore natural to wonder if this latter bracket can be recovered from
that canonically defined on the direct sum $\frak g^n$ through a Lie
algebra deformation or a Lie algebra (Wigner) contraction. This latter is actually 
the case though the contraction which must be considered turns out to be a
little more complicate then those usually considered in the literature
\cite{BR}. Therefore before giving the general construction let us 
describe the two simpler case in order to explain
which kind of problems arise.\par
Actually for the first non trivial case (namely $n=1$) one can still
implement the usual Lie algebra construction up to a linear transformation.
Since ${\frak g}^1$ is the direct sum of two copies of $\frak g$ we can consider 
on it a basis $X^{(k)}_i$ $k=0,1$  $i=1,\dots,r=\dim(\frak g)$ such that 
$X^{(k)}_i\in \frak g^k$ $i=1,\dots,r$ for $k=0,1$ respectively and with
respect to which the Lie bracket canonically defined on $\frak g^1$ are
given by:
\beq
\label{clbn1}
\begin{array}{lll}
\left[ X^{(l)}_i,X^{(l)}_j\right]&=\sum_{k=0}^rc^k_{ij}X^{(l)}_k  &\quad l=0,1
\quad i,j=1,\dots r \\
\left[ X^{(0)}_i,X^{(1)}_j \right]&= 0   & \quad i,j=1 , \dots r
\end{array}
\eeq
i.e., with the same structure constant for each copy of $\frak g$.
Then by  performing  the parameter depending change of basis
$$
\begin{array}{lll}
Y^{(0)}_i=X^{(0)}_i+X^{(1)}_k&\qquad i=1,\dots,r\\
Y^{(1)}_i= \zeta X^{(1)}_i&\qquad i=1,\dots,r
\end{array}
$$
the  equations \rref{clbn1} can be written (in the new basis) as
$$
\begin{array}{lll}
\left[ Y^{(0)}_i,Y^{(0)}_j\right] =&\sum_{k=0}^rc^k_{ij}Y^{(0)}_k&\quad i,j=1,\dots r\\
\left[ Y^{(0)}_i,Y^{(l)}_j\right] =&\sum_{k=0}^rc^k_{ij}Y^{(l)}_k&\quad i,j=1,\dots r\\
\left[ Y^{(l)}_i,Y^{(l)}_j\right] =&\zeta \sum_{k=0}^rc^k_{ij}Y^{(l)}_k&\quad i,j=1,\dots r.
\end{array}
$$
Now by computing the limit $\zeta\to 0$ in these latter equations one
obtains exactly the Lie bracket previously defined on $\frak g^{(1)}$.
Unfortunately this procedure can not be directly and straightforward
extended to the general case. Let us indeed examine   the next case
(i.e., $n=2$). Here we consider again a basis $X^{(k)}_i$ $i=1,\dots,r$ $k=0,1,2$
$X^{(k)}_i\in \frak g^k$ $i=1,\dots,r$ $k=0,1,2$ with respect to which the
canonical Lie bracket can be written in the form
\beq
\label{clbn2}
\begin{array}{lll}
\left[X^{(l)}_i,X^{(l)}_j\right]=&\sum_{k=0}^rc^k_{ij}X^{(l)}_k&\qquad l=0,1,2\quad
i,j=1,\dots r\\
\left[X^{(0)}_i,X^{(1)}_j\right]=& 0&\qquad i,j=1,\dots r
\end{array}
\eeq
(again therefore with the same structure constants for each copy of $\frak
g$). The case $n=2$ seems to suggest to perform first the (parameter
depending) change of basis given by the equation
$$
\begin{array}{lll}
Y^{(0)}_i=X^{(0)}_i+X^{(1)}_i&\qquad i=1,\dots,r\\
Y^{(1)}_i=\zeta X^{(0)}_i+\zeta^2X^{(1)}_i&\qquad i=1,\dots,r
Y^{(2)}_i=\zeta^2 X^{(0)}_i&\qquad i=1,\dots,r
\end{array}
$$
With respect to this new  basis the equations \rref{clbn2} becomes
$$
\begin{array}{lll}
\left[Y^{(0)}_i,Y^{(0)}_j\right]=&\sum_{k=0}^rc^k_{ij}Y^{(0)}_k&\quad i,j=1,\dots r\\
\left[Y^{(1)}_i,Y^{(l)}_j\right]=&\sum_{k=0}^rc^k_{ij}(Y^{(l)}_k+\zeta^2Y^{(1)}_k-\zeta Y^{(2)}_k)
&\quad i,j=1,\dots r\\
\left[Y^{(l)}_i,Y^{(2)}_j\right]=&\zeta \sum_{k=0}^rc^k_{ij}Y^{(2)}_k&\quad i,j=1,\dots r\\
\left[Y^{(2)}_i,Y^{(2)}_j\right]=&\zeta^2 \sum_{k=0}^rc^k_{ij}Y^{(2)}_k&\quad i,j=1,\dots r.
\end{array}
$$
here in the formula for $\left[Y^{(1)}_i,Y^{(l)}_j\right]$ appear ``unwanted terms''
which luckily enough in this specific case disappear when we perform the
limit $\zeta\to 0$ giving namely  the Lie bracket for $\frak g^{(2)}$.
But nevertheless it is clear that to control the behaviour of such
``unwanted terms''  when the parameter $\zeta$ goes to zero will be the
main issue of the general case. Indeed without a carefully choice of the
dependence from the parameter $\zeta$ of the transformation of basis in
$\frak g^n$ one should fear that these terms may not go to zero or even
explode to infinity. Fortunately there exist opportune choices of the
$\zeta$--depending basis transformation such that  holds  the
\begin{prop} \label{liealgcon} Let ${\frak g}^n=\bigoplus_{i=0}^n{\frak g}_i$
$\frak g_i\simeq \frak g$ the direct sum of $n+1$ copies of the Lie
algebra $\frak g$ and let be $\{X^{(l)}_i\}$ $l=0,\dots,n$,
$i=1,\dots,r=\dim(\frak g)$ a basis for ${\frak g}^n$ such that
$X^{(l)}\in \frak g_l$ for all $i=1,\dots,r$ and such that with respect to
this basis the canonical bracket for ${\frak g}^n$ can be written as
\beq
\label{canbraxn}
\begin{array}{lll}
\left[X^{(l)}_i,X^{(l)}_j\right]=&\sum_{k=1}^rc^k_{ij}X^{(l)}_k&\qquad
l=0,\dots,n\quad i,j=1,\dots,r\\
\left[X^{(l)}_i,X^{(m)}_j\right]=&0&\qquad
l,m=0,\dots,n, l\neq m\quad i,j=1,\dots,r.
\end{array}
\eeq
Thus if we write the equations \rref{canbraxn} with respect to the new basis
\beq
\label{Ybgn}
Y^{(l)}_i(\zeta)=\sum_{s=0}^{n-l}\zeta^{l2^s}X^{(s)}_i\quad l=0,\dots,n\quad
i=1,\dots,r
\eeq
and then perform the limit $\zeta\to 0$ we obtain the Lie bracket for
the Lie algebra $\frak g^{(n)}$ \rref{lbmm}:
\beq
\label{eqgnf}
\begin{array}{lll}
\left[Y^{(l)}_i,Y^{(m)}_j\right]=&\sum_{k=1}^rc^k_{ij}Y^{(l+m)}_k&\qquad l+m\leq n
\quad i,j=1,\dots,r\\
\left[Y^{(l)}_i,Y^{(m)}_j\right]=&0&\qquad l+m> n
\quad i,j=1,\dots,r
\end{array}
\eeq
\end{prop}
{\bf Proof } 
Let us first compute Lie bracket of two elements of the new basis say $Y^{(k)}_p$ 
and $Y^{(j)}_q$ (where we omit here and in what follow the explicit dependence from 
$\zeta$) in terms of the old one:
\beq
\label{wigcomm1}
[Y^{(k)}_p,Y^{(j)}_q]=\sum_{t=1}^{{\rm dim} \frak{g}} \sum_{l=0}^{{\rm Min}(n-k,n-j)}
\zeta^{(j+k)2^l}c_{pq}^{t}X_{t}^{(l)}
\eeq
If  $k+j >  n$ we leave the expression \rref{wigcomm1} invariate, otherwise, 
if $k+j \leq  n$, we write it as:
\beq
\label{wigcomm2}
[Y^{(k)}_p,Y^{(j)}_q]=
\sum_{t=1}^{{\rm dim} \frak{g}} \sum_{l=0}^{n-k-j}\zeta^{(j+k)2^l}c_{pq}^{t}X_{t}^{(l)} +
\sum_{t=1}^{{\rm dim} \frak{g}} \sum_{l=n-k-j+1}^{{\rm Min}(n-k,n-j)} 
\zeta^{(j+k)2^l}c_{pq}^{t}X_{t}^{(l)}
\eeq
which, using the definition of the elements $Y^{(k)}_p $, can be written as
\beq
\label{wigcomm3}
[Y^{(k)}_p,Y^{(j)}_q]=
\sum_{t=1}^{{\rm dim} \frak{g}}c_{pq}^{t}Y_{t}^{(l)}+
\sum_{t=1}^{{\rm dim} \frak{g}} \sum_{l=n-k-j+1}^{{\rm Min}(n-k,n-j)} 
\zeta^{(j+k)2^l}c_{pq}^{t}X_{t}^{(l)}.
\eeq
What we have to show is that if we perform the limit for $\zeta \to 0$
then the whole right hand of equation \rref{wigcomm1}
and the second term in the right hand of   \rref{wigcomm3} vanishes. 
Now  the \rref{wigcomm1}  can be 
written as:
\beq
[Y^{(k)}_p,Y^{(j)}_q]=\sum_{t=1}^{{\rm dim} \frak{g}} 
\zeta^{{\rm Min}(j,k) 2^l}c_{pq}^{t}Y^{{\rm Max}(k,j)}_{t}
\eeq
and, since $k+j>n$ with $k,j \leq n$, implies ${\rm Min}(j,k) \geq 1 $ performing the
limit for  $\zeta \to 0$ we get 
\beq
[Y^{(k)}_p,Y^{(j)}_q]=0.
\nonumber
\eeq
In the case of equation \rref{wigcomm3}, i.e. $k+j \leq n$, in order to avoid to invert 
completely equations \rref{Ybgn} 
let us prove the following 
\begin{lem}
The elements $2^{k(n-k)} X_p^{(k)}$ expressed in the basis $Y^{(n)}_j$ have the form:
\beq
\label{XintermsofY}
\zeta^{2^k(n-k)}X^{(k)}_p=\sum_{j=0}^{k}q_j^k(\zeta^{-1})Y_p^{(n-j)}
\eeq
where $ {\rm deg}(q^k_0)=1$ and ${\rm deg} (q_j^k)=2^k-2^j$
\end{lem}
{\bf Proof }
We proceed by induction. The result for $k=0$ is immediately true for $k=1$ we have 
$\zeta^{2(n-1)}X^{(1)}=Y^{(n-1)}-\zeta^{-1}Y^{(n)}_p$.\\
So let us suppose \rref{XintermsofY} true for $k$ and let us check it for $k+1$.\\
We have 
$$
Y^{(n-k-1)}_p=\sum_{l=0}^k \zeta^{2^l(n-k-1)}X_p^{(l)}+\zeta^{2^{k+1}(n-k-1)}X_p^{(k+1)}
$$
which can be written as 
$$
\zeta^{2^{k+1}(n-k-1)}X^{(n+1)}_p=Y^{(n-k-1)}_p-\sum_{l=0}^k \zeta^{2^l(l-k-1)}\zeta^{2^l(n-l)}
X^{(l)}_p,
$$
then by induction
$$
\zeta^{2^{k+1}(n-k-1)}X^{(n+1)}_p=Y^{(n-k-1)}_p-
\sum_{l=0}^k \sum_{j=0}^l\zeta^{2^l(l-k-1)}q_j^l(\zeta^{-1})Y^{(n-j)}_p.
$$
If we commute the two sum in this last expression 
$$
\zeta^{2^{k+1}(n-k-1)}X^{(n+1)}_p=Y^{(n-k-1)}_p-\sum_{j=0}^k 
\left( \sum_{l=j}^k\zeta^{2^l(l-k-1)}q_j^l(\zeta^{-1}) \right) Y^{(n-j)}_p.
$$
From which we have 
$$
q_j^{k+1}(\zeta^{-1})=-\sum_{l=j}^k\zeta^{2^l(l-k-1)}q_j^l(\zeta^{-1})
$$
where ${\rm deg}(q_0^k)=1$ and
$$
{\rm deg} (q_j^{k+1})=2^k+{\rm deg} (q_j^{k})=2^k+2^k-2^j=2^{k+1}-2^j.
$$
This concludes the proof of the lemma.
\endpf 
Now using this  lemma we can estimate $\zeta^{2^l(j+k)}X_p^{(l)}$, we have indeed that 
\bean
\zeta^{2^l(j+k)}X_p^{(l)}&=&\zeta^{2^{l}(j+k)+2^{l}(n-l-n+l)}X_p^{(l)}
=\zeta^{2^{l}(j+k-n+l)}\zeta^{2^{l}(n-l)}X_p^{(l)}\\
&=  &\zeta^{2^{l}(j+k-n+l)}\sum_{j=0}^{l}q_j^k(\zeta^{-1})Y^{(n-j)}
\eean
Therefore since ${\rm deg}( q_j^l) \leq 2^{l}-1 $ and $(j+k-n+l)>1$:
$$
\lim_{\zeta \to 0} \zeta^{2^{l}(j+k-n+l)}q_j^k(\zeta^{-1})=0 \qquad \forall l \quad n-k-j+1 \leq l \leq 
{\rm Min}(n-k,n-j).
$$
\endpf
It is natural to wonder if our generalised Wigner's contraction 
allows us to obtain also the wanted 
ad--invariant symmetric non degenerate bilinear forms on $\frak g^{(n)}$ 
\rref{omlkbf} as well. This result can be actually achieved but unfortunately one has first
to modify the canonical bilinear form defined on the direct sum 
${\frak g}^n=\oplus_{k=0}^n {\frak g}$ by multiplying its entries by factors depending
in a quite complicate way from the parameter $\zeta$. More precisely it holds
the
\begin{prop}\label{wrblf} Let us consider on the Lie algebra given by the 
direct sum ${\frak g}^n=\oplus_{k=0}^n {\frak g}$ the canonical bilinear form induced
from that defined on $\frak g$ i.e. in formula if 
$X=(X_0,\dots,X_n)$ and $Y=(Y_0,\dots,Y_n)$ are two elements of ${\frak g}^n$ 
then 
\beq
\label{ninbf}
\langle X,Y\rangle^n=\sum_{k=0}^n\langle X_k,Y_k\rangle_{\frak g} 
\eeq 
where $\langle \cdot,\cdot \rangle_{\frak g}$ denotes the  symmetric bilinear form defined 
on $\frak g$. Then there exist $\zeta$--depending 
factors $d_k(\zeta)$ such  that the modified inner product on $\frak g^n$ given 
by 
\beq
\label{mdip}
\langle X,Y\rangle^n(\zeta)=\sum_{k=0}^nd_k(\zeta)\langle X_k,Y_k\rangle_{\frak g}
\eeq
 has the following property
\beq
\label{lwblf}
\mbox{lim}_{\zeta\to 0} \langle Y^{(j)}_p(\zeta),Y^{(k)}_q(\zeta)\rangle^n(\zeta)= 
 \langle Y^{(j)}_p,Y^{(k)}_q\rangle^{(n)}_{\cal A}  
\eeq
\end{prop}
{\bf Proof.}
Let us first consider the relations \rref{lwblf} when $j=0$ and  $k$ which running
between $0$ and $n$. This relations form a system of $n+1$ independent linear 
equations
\beq
\label{synp1}
\langle Y^{(0)}_p(\zeta), Y^{(k)}_q(\zeta)\rangle =a_k\omega_{pq}\qquad k=0,\dots n,\ p,q=1,\dots \dim(\frak g)
\eeq
where $\{\omega_{pq}\}_{p,q =1,\dots,\dim{\frak g}}$ is the matricial form of the pairing 
$\langle \cdot,\cdot\rangle_{\frak g}$.\par\noindent
We actually should simply verify the weaker assumption
$$ 
\langle Y^{(0)}_p(\zeta), Y^{(k)}_q(\zeta)\rangle \sim a_k\omega_{pq}\quad  k=0,\dots n,\;  p,q=1,\dots,\dim(\frak g)
$$
but since this does not affect our 
proof, let us consider the equation \rref{synp1} instead. 
From this equation one can explicitly compute the coefficients $d_k(\zeta)$, $k=0,\dots,n$,
altought their expression in terms of $\zeta$ turns out to be complicate. Lakely enough
we need only to know that the $d_k(\zeta)$  satisfy \rref{synp1} and use it to
to prove the following technical
\begin{lem}\label{aexal} The coefficients $d_k(\zeta)$ have the following 
asymptotic  expansion
\beq
\label{aexa}
{\zeta}^{(n-k)2^{k}}d_k(\zeta)\sim (-1)^k a_n\zeta^{-2^k+1}.
\eeq
\end{lem}
{\bf Proof.} Let us proceed by induction from formula \rref{synp1}  
we have immediately
$$
\zeta^{n}d_0(\zeta)=a_n
$$
and similarly
$$
\zeta^{(n-1)2}d_1(\zeta)\sim - a_n \zeta^{-2+1}.
$$
Let us suppose \rref{aexa} true for $k$ and let us prove it for $k+1$.
Using \rref{synp1}  we have
$$
\zeta^{(n-k-1)2^{k+1}}d_{k+1}(\zeta)=a_{n-k-1}-\sum_{i=0}^k \zeta^{(n-k-1)2^i}d_i(\zeta)
$$
therefore using the induction hypothesis we obtain
$$
\begin{array}{ll}
\zeta^{(n-k-1)2^{k+1}}d_{k+1} (\zeta)=&a_{n-k-1}-\sum_{i=0}^k \zeta^{(i-k-1)2^i}\zeta^{2^i(n-i)}d_i(\zeta)  \\
& \sim a_{n-k-1} -\sum_{i=0}^k (-1)^i\zeta^{(i-k-1)2^i-2^i+1}a_n  \\
 & \sim -\sum_{i=0}^k (-1)^i a_n \zeta^{2^i(i-k-2)+1}\sim (-1)^{k+1}a_n \zeta^{-2^{k+1}+1}
\end{array}
$$
which is formula \rref{aexa} for $k+1$ proving the claim.
\endpf
We can complete our proposition.
For $j+k> n$ we have indeed
\beq
\label{yjkpn}
\langle Y^{(j)}_p(\zeta),Y^{(k)}_q(\zeta)\rangle(\zeta)=\sum_{l=0}^{\rm{Min}\{n-k,n-j\}}
d_l(\zeta)\zeta^{(j+k)2^l}\langle X_p^{(l)},X_q^{(l)}\rangle_{\frak g}.
\eeq
Using lemma \ref{aexal} we can estimate this equation as 
$$
\begin{array}{ll}
 & \sum_{l=0}^{\rm{Min}\{n-k,n-j\} }{\zeta}^{(j+k-n+l)2^l}\zeta^{(n-l)2^l}d_l(\zeta)\\  
 &\sim  \sum_{l=0}^{\rm{Min}\{n-k,n-j\} }(-1)^la_n{\zeta}^{(j+k-n+l)2^l-2^l+1}=
  \sum_{l=0}^{\rm{Min}\{n-k,n-j\} }(-1)^la_n{\zeta}^{2^l(j+k-n+l-1)+1}
\end{array}
$$
and since $j+k-n+l> 0$ (being $l\geq 0$) we have that 
$$
\mbox{lim}_{\zeta\to 0}\langle Y^{(k)}_p,Y^{(j)}_q\rangle=0 \qquad \mbox{for\ } j+k> n
$$
as wanted. If viceversa $j+k\leq n$, we have that 
as 
$$
\begin{array}{ll}
\langle Y^{(k)}_p,Y^{(j)}_q\rangle &=\sum_{l=0}^{\rm{Min}\{n-k,n-j\} }{\zeta}^{(j+k)2^l}d_l(\zeta)\omega_{pq}\\
&=\sum_{l=0}^{n-k-j}{\zeta}^{(j+k)2^l}d_l(\zeta)\omega_{pq}+
\sum_{l=n-k-j+1}^{\rm{Min}\{n-k,n-j\} }{\zeta}^{(j+k)2^l}d_l(\zeta)\omega_{pq}
\end{array}
$$
using \rref{synp1}   we obtain immediately that 
$$
\sum_{l=0}^{n-k-j }{\zeta}^{(j+k)2^l}d_l(\zeta)=a_{k+j},
$$
while for the second summand we have, using lemma \ref{aexal}
$$
\sum_{l=n-k-j+1}^{\rm{Min}\{n-k,n-j\} }{\zeta}^{(j+k)2^l}d_l(\zeta)\sim 
\sum_{l=n-k-j+1}^{\rm{Min}\{n-k,n-j\} }a_n {\zeta}^{(j+k-n+l-1)2^l+1}.
$$
But this implies, because $l> n-k-j$ and therefore $j+k-n+l-1\geq 0$
that the parameter $\zeta$ in the addends of  the above written equation appears 
with powers bigger then one, and therefore that 
$$
\mbox{lim}_{\zeta\to 0}\left( \sum_{l=n-k-j+1}^{\rm{Min}\{n-k,n-j\} }{\zeta}^{(j+k)2^l}
d_l(\zeta) \right) =0
$$
and then finally that
$$
\mbox{lim}_{\zeta\to 0}\langle Y^{(k)}_p,Y^{(j)}_q\rangle=\omega_{pq} \qquad \mbox{for\ } j+k\leq n
\quad j,k\geq 1\; p,q=1,\dots,\dim(\frak g)
$$
which concludes the proof of our proposition.
\endpf
Unfortunately both the constructions of $\frak{g}^{(n)}$ so far considered
are not particularly  well suited for the purposes we have in mind, asking
for a more handable one.\\
Therefore the last part of this section is devoted to tackle this problem
and to present a matricial realization of $\frak{g}^{(n)}$.\\
The main result in order to achieve our task is the following
\begin{theorem}
\label{isomorph}
For every $n$ and $\frak{g}$ there is an isomorphism $\Phi$ 
between the polynomial Lie algebra $\frak{g}^{(n)}$ 
and  the tensor product $\frak{g} \otimes {\Bbb C}^{(n)}(\lambda)$:
where $ {\Bbb C}^{(n)}(\lambda)$ is the commutative quotient algebra 
$ {\Bbb C}(\lambda)/{\cal I}^{(n)}(\lambda)$ defined in \ref{quotspc}
\end{theorem}
{\bf Proof}
Let $\Phi$ be the linear map:
$$
\Phi: \frak{g}^{(n)} \longrightarrow \frak{g} \otimes {\Bbb C}^{(n)}(\lambda)
$$
defined by
$$
\Phi(X_0, \dots , X_n)=\sum_{i=0}^{n}X_i \otimes \lambda^i.
$$
This map is obviously a Lie algebra homomorphism, we have indeed:
\bean
\Phi([(X_0,\dots,X_n),(Y_0,\dots,Y_n)])&=&\Phi((Z_0,\dots,Z_n))
=\sum_{i=0}^{n}Z_i \otimes \lambda^i\\
= \sum_{i=0}^{n} \sum_{j=0}^{i}[X_j,Y_{i-j}] \otimes \lambda^i
&=&  \left[\sum_{k=0}^n X_k \otimes \lambda^k ,\sum_{j=0}^n Y_j\otimes \lambda^j\right]\\  
&=&[\Phi((X_0,\dots,X_n)),\Phi((Y_0,\dots,Y_n))].
\eean
Since ${\rm dim} (\frak{g}^{(n)})=(n+1){\rm dim}(\frak{g})$ we have only to check
that ${\rm Ker}(\Phi)=0$ but this follows immediately from the definition of $\Phi$.
\endpf
On behalf of the previous theorem, it remains only  
to look for a true matricial representation Lie algebra ${\Bbb C}^{(n)}(\lambda)$.
Let us therefore prove the 
\begin{prop}
\label{homomring}
The map $\rho$ given by
\beq
\label{rhcl}
\begin{array}{ll}
\rho:{\Bbb C}^{(n)}(\Lambda) \longrightarrow& {\rm End}({\Bbb C}^{(n+1)}) \\
\rho(c_i \otimes \Lambda^i)\mapsto& c_i \lambda^i
\end{array}
\eeq
where $\Lambda$ is the $(n+1) \times (n+1)$ matrix given by
\beq
\label{efol}
\Lambda=\sum_{i=0}^{n} e_{i+1,i} 
\eeq
and 
$$
(e_{ij})_{kr}=\left\{ \begin{array}{ll}
                         1 & \mbox{if $i=j$, $k=r$} \\
                         0 & \mbox{otherwise}
                       \end{array}              
                \right.
$$
is a ring homomorphism. 
\end{prop}
{\bf Proof }
We have indeed that
$$
\rho(\Lambda^i)\rho(\Lambda^j)=\Lambda^{i+j}=\left\{ \begin{array}{ll}
                                                     \rho(\Lambda^{i+j}) & i+j\leq k \\
                                                      0 & \mbox{otherwise}. 
                                                     \end{array} \right.
$$
\endpf
Now using together Theorem \ref{isomorph} and and proposition 
\ref{homomring} we get a matricial representation of $\frak{g}^{(k)}$.
\begin{theorem}
\label{reprtens}
If $\Pi: \frak{g} \longrightarrow {\rm Aut}({\Bbb C}^{m})$ for some m is a true
representation of $\frak{g}$ then the map 
$$
\widetilde{\Pi}: \frak{g}^{(n)} \mapsto {\rm Aut}({\Bbb C}^{m(n+1)})
$$
given by
{\small \bea \label{reptenseq}
\widetilde{\Pi}(X_0, \dots ,X_n)&=& \sum_{i=0}^n X_i \Lambda^i \\
&=&\left( \begin{array}{cccccc}
                      \Pi(X_0)& 0& \quad 0 \quad& \quad 0 \quad& 0& 0 \\
                      \Pi(X_1)&\Pi(X_0)& 0 &0&0& 0 \\
                      \vdots &\ddots&\ddots & \ddots& 0& 0 \\
                      \vdots  &\vdots &\ddots &\ddots & \ddots&\vdots \\
                      \Pi(X_{n-1}) &\vdots  & \dots & \dots &\Pi(X_0)  & 0 \\
                      \Pi(X_{n})&\Pi(X_{n-1}) &\dots &\dots &\Pi(X_1) &\Pi(X_0) 
        \end{array} \right) \nonumber 
\eea}
is a true representation of $\frak{g}^{(n)}$.
\end{theorem}
{\bf Proof }
Since we have constructed a representation of ${\Bbb C}(\lambda^n)$ we have a representation
of $\frak{g} \otimes {\Bbb C}(\lambda^n)$ given by:
$$
\Pi \otimes \rho : \frak{g} \otimes {\Bbb C}(\lambda^{n})
\longrightarrow {\rm End}({\Bbb C}^{m}) \otimes {\rm End}({\Bbb C}^{n+1}) 
\cong {\rm End}({\Bbb C}^{m(n+1)}) 
$$
To bring it on $\frak{g}^{(n)}$ directly we have only to use the isomorphism 
$\Phi:\frak{g}^{(k)} \cong \frak{g} \otimes {\Bbb C}(\lambda^{k}) $:
$$
\widetilde{\Pi}=\Pi \circ \rho  \circ \Phi:\frak{g}^{(k)} \longrightarrow {\rm Aut}({\Bbb C}^{m(n+1)})
$$ 
\endpf
Previously in this section we have shown that if $\frak{g}$ possesses
an ad--invariant bilinear non degenerate  form this gives rise a 
 ad--invariant bilinear non degenerate form on $\frak{g}^{(n)}$. 
It is therefore 
natural to wonder if this latter form has a natural expression in our matricial
representation. This is actually the case. We have indeed for instance when $a_k=1$ for all 
$k$:
\beq
\label{matformbl}
\langle (X_0,\dots,X_n),(Y_0,\dots,Y_n)\rangle^{(n)}=\rm{tr}(\widetilde{\Pi}((X_0,\dots,X_n)
\widetilde{\Pi}((Y_0,\dots,Y_n)C^{(n)})
\eeq
where $C^{(n)}$ is the $m(n+1)\times m(n+1)$ matrix:
$$
C^{(n)}=\left(\begin{array}{cccccc} \frac{1}{n+1}{\Bbb I}_m & \frac1n{\Bbb I}_m& \dots & 
\frac13 {\Bbb I}_m & \frac12 {\Bbb I}_m  & {\Bbb I}_m \\
 0 & \frac{1}{n+1}{\Bbb I}_m & \frac1n {\Bbb I}_m  & \dots & \frac13 {\Bbb I}_m  & 
\frac12 {\Bbb I}_m \\
0 &0 & \ddots & \ddots & \ddots & \frac13 {\Bbb I}_m \\
\vdots &\vdots   & 0 & \ddots & \ddots & \vdots\\
\vdots &\vdots  & \vdots & 0 & \frac{1}{n+1}{\Bbb I}_m   & \frac1n {\Bbb I}_m \\
0 &0   & 0  & \dots & 0 & \frac{1}{n+1}{\Bbb I}_m \end{array}\right)
$$
i.e.  $C^{(n)}_{p,p+k}=\frac{1}{n+1-k}{\Bbb I}_m$, $p=0,\dots n-k$ and
$k=0,\dots,n$   and  $C^{(n)}_{pq}=0$ if $q<p$, where ${\Bbb I}_m$ denotes the $m\times m$ identity
matrix.
\section{The affine Lie algebras}
In the previous section we have constructed a class of non semisimple Lie 
algebras which posses an ad--invariant non degenerate symmetric bilinear form.
This peculiar property suggests to investigate their affinization. To this 
task is devoted this third section. For the construction of the affine Lie 
algebras  we will follows that presented by Kac in his famous book \cite{K},
with the only difference that we will end up with multidimensional central 
extensions, namely the affinization of the Lie algebra ${\frak g}^{(n)}$ will
have $n+1$ central charges. \par
Let us  consider a polynomial Lie algebra ${\frak g}^{(n)}$ 
where $\frak g$ is  semisimple and let us first define the 
corresponding  loop algebra 
\beq
\label{lgka}
{\cal L}({\frak g}^{(n)})={\frak g}^{(n)}\otimes_{\Bbb C} {\Bbb C}(t,t^{-1})
\eeq
where ${\Bbb C}(t,t^{-1})$ denotes as usual  the algebra of Laurent 
polynomials in a complex variable $t$. On it is defined an infinite 
complex Lie algebra bracket
$[\cdot,\cdot]_0$ by
$$
[X\otimes p,Y\otimes q]_0=[X,Y]\otimes pq \qquad (p,q\in {\Bbb C}(t,t^{-1});
 X,X \in {\frak g}^{(n)}).
$$
Now our non degenerated ad--invariant symmetric bilinear form 
$\langle \cdot, \cdot\rangle_{\cal A}^{(n)}$ \rref{glbilin} can be extended to a
${\Bbb C}(t,t^{-1})$--valued bilinear form $\langle \cdot,\cdot\rangle^{(n)t}_{\cal A}$
on ${\cal L}({\frak g}^{(n)})$ by
\beq
\label{blfol}
\langle X\otimes p, Y\otimes q\rangle^{(n)}_{\cal A}(t)=
\langle X, Y\rangle^{(n)}_{\cal A}pq.
\eeq
Moreover this latter expression can be used to define a symmetric  
non degenerated ad--invariant bilinear form on  ${\cal L}({\frak g}^{(n)})$
as 
\beq
\label{saindf}
\langle X\otimes p, Y\otimes q\rangle^{(n)t}_{\cal A}=
\mbox{Res}(\langle X, Y\rangle^{(n)}_{\cal A}pq)
\eeq
where the $\mbox{Res}$ is the linear functional of ${\Bbb C}(t,t^{-1})$ 
defined by the properties 
$$
\mbox{Res}(t^{-1})=1; \qquad \mbox{Res}(\frac{dp}{dt}=0). 
$$
The key point in the possessing of a non degenerated  
ad--invariant symmetric bilinear form is that it allows us to define a
$\Bbb C$--valued 2--cocycle on the Lie algebra ${\frak g}^{(n)}$ 
$\Psi: {\cal L}({\frak g}^{(n)}) \to \Bbb C$ (see \cite{K})  as 
\beq
\label{2cocy}
\Psi(X(t),Y(t))=\mbox{Res}(\langle \frac{dX(t)}{dt}, Y(t)\rangle^{(n)}_{\cal A})(t).
\eeq
This 2--cocycle in turns allows us to extend our Lie algebra 
${\cal L}({\frak g}^{(n)})$ by a $n+1$--dimensional center. Explicitly
\beq
\label{ltd} 
\tilde{{\cal L}}({\frak g}^{(n)})={\cal L}({\frak g}^{(n)})\oplus\sum_{i=0}^n
\oplus {\Bbb C}c_i
\eeq
with Lie bracket defined as
\beq {\footnotesize 
\label{lieca}
\begin{array}{ll}
[(X_0,\dots,X_n)\otimes p,(Y_0,\dots,Y_n)\otimes q]&=
[(X_0,\dots,X_n)\otimes p,(Y_0,\dots,Y_n)\otimes q]_0\\
&+\sum_{i=0}^n(\sum_{j=0}^i\Psi(X_{i-j}\otimes p,Y_j\otimes q)c_j.
\end{array} }
\eeq 
Again, as usual, we also extend every derivation $D$ of the Lie algebra
${\Bbb C}(t,t^{-1})$ to a derivation of the whole Lie algebra 
${\cal L}({\frak g}^{(n)})$ by setting
\beq
\label{derla}
D(X\otimes p)=X\otimes D(p).
\eeq
In particular we denote by $\widehat{{\cal L}}({\frak g}^{(n)})$ 
the Lie algebra obtained
by adding to ${\cal L}({\frak g}^{(n)})$ a derivation $d$ which acts on 
${\frak g}^{(n)}$ as $t\frac{d}{dt}$ and which annihilates the central charges
$c_i$ $i=0,\dots,n$. 
Thus we can extend the bilinear form \rref{blfol} 
to a symmetric ad--invariant bilinear form 
on the whole algebra ${\cal L}({\frak  g}^{(n)})$ by setting 
$\langle c_i, d\rangle^{(n)t}_{\cal A}=1$,
$\langle c_i, c_j\rangle^{(n)t}_{\cal A}=\langle d,d\rangle^{(n)t}_{\cal A}=0$ for $i,j=0,\dots n$. 
Then explicitly $\widehat{{\cal L}}({\frak g}^{(n)})$
 is the vector space 
\beq
\label{lhvs}
\widehat{{\cal L}}({\frak g}^{(n)})={\cal L}({\frak g}^{(n)})\oplus\sum_{i=0}^n
\oplus {\Bbb C}c_i\oplus{\Bbb C}d
\eeq
with bracket defined as follows 
\beq
{\footnotesize
\label{lbohgn}
\begin{array}{ll}
 &[(X_0,\dots,X_n)\otimes t^p\oplus(\sum_{i=0}^n \nu_i c_i\oplus \mu d),
(Y_0,\dots,Y_n)\otimes t^q\oplus(\sum_{i=0}^n \nu^1_i c_i\oplus \mu^1 d)]=\\
&[(X_0,\dots,X_n),(Y_0,\dots,Y_n)]\otimes t^{p+q}+(\mu q(Y_0,\dots,Y_n)
\otimes t^q-\mu^1 p(X_0,\dots,X_n)\otimes t^p)\\
&+ p\delta_{p,-q}
\sum_{i=0}^n\sum_{j=0}^ia_j\langle X_{i-j},Y_{j}\rangle c_i. 
\end{array} }
\eeq
The algebra $\widehat{{\cal L}}({\frak g}^{(n)})$ defined above is the 
``affine  Lie  algebra'' which was looking for. 
It should be here pointed out that this algebra could be also obtained as 
quotient from the Lie algebra 
$\widehat{{\cal L}}({\frak g})\otimes {\Bbb C}(\lambda)$, where 
$\widehat{{\cal L}}({\frak g})$ is the usual affine non twisted Lie algebra  
associated with  the semisimple Lie algebra $\frak g$, using the same procedure 
which gives the Lie algebra ${\frak g}^{(n)}$ starting 
from the Lie algebra $\frak g\otimes {\Bbb C}(\lambda)$. 
Moreover is still true that there exists a isomorphism between the Lie algebras
$\widehat{{\cal L}}({\frak g}^{(n)})$ and 
$\widehat{{\cal L}}({\frak g})\otimes {\Bbb C}^{(n)}(\lambda)$ where 
${\Bbb C}^{(n)}(\lambda)$ is the commutative nilpotent Lie algebra defined 
in \rref{quotspc}, which is the obvious extension of the corresponding isomorphism
for the finite dimensional Lie algebra ${\frak g}^{(n)}$ defined in \ref{pla1d}.
This isomorphism explains very clearly the phenomenon of the multidimensional
central extension, this subspace coincides indeed with the $N$--dimensional 
tensor product $c\otimes {\Bbb C}^{(n)}(\lambda)$ where $c$ is the unique 
central charge of $\widehat{{\cal L}}({\frak g})$.\par 
Since  the aim of the next section is the construction of the vertex operators 
of these algebra, which  roughly speaking represent certain generating 
series of elements of $\widehat{\cal L}({\frak g}^{(n)})$ rather then 
individual ones, it is opportune to cast formulas \rref{lbohgn}
in a more compact form using the formal calculus (see for example \cite{FLM} 
or \cite{K1} for more details). \par
The first step in this direction  is to consider the Cartan 
decomposition of our semisimple complex Lie algebra ${\frak g}$ 
\beq
\frak g={\frak h}\oplus\sum_{\alpha\in \Delta}\oplus{\frak g}_{\alpha},
\label{card}
\eeq
where $\frak h$ is an once for ever fixed  Cartan subalgebra of ${\frak g}$, 
and $\Delta=({\frak g},{\frak h})$ is the corresponding the root system. Let 
us denote
further by $\Sigma=\{\alpha_1,\dots\alpha_r\}$ a subset of simple roots in 
$\Delta$  with $r=\dim(\frak h)=\mbox{rank}(\frak g)$ (again once for ever 
fixed) and let $\{H_{\alpha_1},\dots, H_{\alpha_r}\}$ be the corresponding set 
of coroots in $\frak h$. Then it is  well known \cite{Hu} 
that $\dim(\frak g)_{\alpha}=1$  $\forall \alpha\in \Delta$ and  that there 
exist non trivial element $X_{\alpha}$ in $g_{\alpha}$ such that 
\beq
\label{bgs}
\{H_{\alpha_1},\dots, H_{\alpha_r}\}\cup\{X_{\alpha}\}_{\alpha\in \Delta}
\eeq
is a basis for $\frak g$, usually called in the literature  Cartan
basis \cite{Hu} and \cite{K}, with
$$
H_{\alpha_i}=[X_{\alpha_i},X_{-\alpha_i}]\qquad i=1,\dots,r.
$$ 
This basis in turn allows us to define a basis 
for the whole Lie algebra $\frak g^{(n)}$ and its affinization 
$\widehat{{\cal L}}({\frak g}^{(n)})$. They will be namely respectively for 
 $\frak g^{(n)}$
\beq
\label{bgsn}
\{H^k_{\alpha_1},\dots, H^k_{\alpha_r}\}\cup\{X^k_{\alpha}\}_{\alpha\in \Delta}
\qquad k=0,\dots n
\eeq
and for $\widehat{{\cal L}}({\frak g}^{(n)})$
 \beq \left\{
\label{bgsna}
\begin{array}{lll}
&\{H^k_{\alpha_1}\otimes t^{m_1},\dots, H^k_{\alpha_r}\otimes t^{m_r}
\}\cup\{X^k_{\alpha}\otimes t^{m_\alpha}\}_{\alpha\in \Delta}
& k=0,\dots n\\
& & m_i\in \Bbb Z \\
& c_0,\dots c_n & \\
& d & 
\end{array} \right.
\eeq
The corresponding Lie bracket are for ${\frak g}^{(n)}$
\beq
\label{lbbgsn}
\begin{array}{ll}
\left[H^k_{\alpha_i},H^j_{\alpha_s}\right]&=0\\
\left[H^k_{\alpha_i},X^j_\alpha\right]&=\left\{ \begin{array}{ll} &
\alpha(H^k_{\alpha_i})
X^{k+j}_{\alpha}\quad  \mbox{if $j+k\leq n$}\\
&  \mbox{ 0 otherwise}\end{array}\right.\\
\left[X^k_{\alpha},X^j_{\beta}\right]&=\left\{ \begin{array}{ll}
& N(\alpha,\beta)X^{k+j}_{\alpha+\beta}  
\mbox{ if $j+k\leq n$ and $\alpha+\beta\in \Delta$}\\
&  \mbox{ 0 otherwise}\end{array}\right.
\end{array}
\eeq
with  opportune integer numbers $N(\alpha,\beta)$.
While for the Lie affine algebra $\widehat{{\cal L}}({\frak g}^{(n)})$ they are
 \beq {\small
\label{afflbbgsn}
\begin{array}{ll}
\left[H^k_{\alpha_i}\otimes t^{m_i},H^j_{\alpha_s}\otimes t^{m_s}\right]&=\langle\
H^k_{\alpha_i},H^j_{\alpha_s}\rangle^{(n)}_{\cal A}\delta_{-m_i,m_s}c_{k+j} \\
\left[H^k_{\alpha_i}\otimes t^{m_i},X^j_\alpha\otimes t^{m_\alpha}\right]&=\left\{ \begin{array}{ll}
& \alpha(H^k_{\alpha_i})
X^{k+j}_{\alpha}\otimes t^{m_i+m_\alpha}\quad  \mbox{if $j+k\leq n$}\\
&  \mbox{ 0 otherwise}\end{array}\right.\\
\left[X^k_{\alpha}\otimes t^{m_\alpha},X^j_{\beta}\otimes t^{m_\beta}\right]&=\left\{ \begin{array}{ll}
N(\alpha,\beta)X^{k+j}_{\alpha+\beta}\otimes t^{m_\alpha+m_\beta} +\langle X^k_{\alpha}, 
X^j_{\beta}\rangle^{(n)}_{\cal A} \delta_{m_\alpha,-m_\beta}c_{j+k}&\\
\qquad  \mbox{if $j+k\leq n$ and $\alpha+\beta\in \Delta$}&\\
  \mbox{ 0 otherwise}&\end{array}\right.\\
\left[d,X\otimes t^m\right]&=m X^k\otimes t^m\qquad \forall X \in \frak g^{(n)}\\
\left[c_i,X\right]&=0\qquad \forall X\in \widehat{{\cal L}}({\frak g}^{(n)}),
 j=0,\dots ,n.
\end{array} }
\eeq
Now we are in the position to introduce our generating series 
 (\cite{K},\cite{FLM}): 
\bea
\label{foce}
\begin{array}{cl}
H^{k}_{\alpha_i}(z)&=\sum_{n \in {\Bbb Z}} H^{k}_{\alpha_i} \otimes t^{n} z^{-n} \\
X^{k}_{\alpha_i}(z)&=\sum_{n \in {\Bbb Z}} X^{k}_{\alpha_i} \otimes t^{n} z^{-n} \\
D_z&=z\frac{\rm{d}}{\rm{d}z}
\end{array}
\eea
where $z$ is a formal variable. This formal operators allow us to cast 
\rref{afflbbgsn} in the 
\begin{lem}
\label{formallbbgsn}
The Lie brackets \rref{lbbgsn} for the affine algebra 
$\widehat{{\cal L}}({\frak g}^{(n)})$
are equivalent to:
\bean
\left[ H^{k}_{\alpha_i}(z_1),H^{j}_{\alpha_s}(z_2) \right] 
            &=&-\langle H^{k}_{\alpha_i}, H^{j}_{\alpha_s}\rangle_{\cal A}^{(n)} 
(D_{z_1}\delta)(z_1/z_2)c_{k+j} \\
\left[ H^{k}_{\alpha_i}(z_1),X^{j}_{\alpha}(z_2) \right] &=& \left\{ \ba{cc}
                              \alpha(H^{k}_{\alpha_i})X^{k+j}_{\alpha}(z_2) \delta(z_1/z_2)&  
k+j \leq n \\
                                               0        &\ {\rm otherwise}
                                                 \ea \right. \\
\left[ X^{k}_{\alpha}(z_1),X^{j}_{\beta}(z_2) \right] &=& \left\{
    \ba{cc} N(\alpha,\beta) X^{k+j}_{\alpha + \beta}(z_2) \delta(z_1/z_2) = & \quad \\  
- \langle X^k_{\alpha}, X^j_{\beta}\rangle_{\cal A}^{(n)} (D_{z_1}\delta)(z_1/z_2)c_{k+j} &  k+j \leq n \\
                                                0        &\ {\rm otherwise}
                                                             \ea \right. \\
\left[d,X(z)\right]&=&-D_{z} X(z) \qquad \forall X \in {\frak g}^{(n)} \\
\left[c_i, X(z) \right] &=& 0 \qquad \forall X(z) \in 
\widehat{{\cal L}}({\frak g}^{(n)}), \quad i=0,\dots ,n
\eean
\end{lem}
{\bf Proof }
The statement follows immediately by  comparing the terms with same degree 
in the equations.
\endpf
Until now we did not impose any restriction on the subset of complex numbers 
$\{a_0,\dots,a_n\}$, which appear in the definition of the bilinear form 
\rref{omlkbf}, but in view of their realization as vertex operator algebra on 
``generalised'' Fock spaces we need to suppose that every $a_k$ is different 
from zero which without loss of generality boils down to set $a_k=1$ for every $k$. 
\section{Vertex algebras representations}
Now we can describe the construction of the  vertex operators 
representation of our Lie algebras $\widehat{{\cal L}}({\frak g}^{(n)})$ in 
the case when $\frak g$ is a simple complex Lie algebra. 
For what observed in the previous sections we can view this generalised 
affine Lie algebra simply as the tensor product:
\beq
\label{ltegcn}
\widehat{{\cal L}}({\frak g}^{(n)})\simeq
\widehat{{\cal L}}({\frak g})\otimes{\Bbb C}^{(n)}
(\lambda).
\eeq
This equivalence suggests of course  a way to obtain a generalised vertex 
operators
representation of $\widehat{{\cal L}}({\frak g}^{(n)})$, namely if 
$\Gamma:\widehat{{\cal L}}({\frak g})\to \rm{End}(V)$ is a vertex operator
 representation for 
$\widehat{{\cal L}}({\frak g})$ and $\rho:{\Bbb C}^{(n)}(\lambda)\to \rm{End}({\Bbb C}^{n+1})$ is 
the representation \rref{reptenseq} of ${\Bbb C}^{(n)}(\lambda)$  
then our ``vertex operators representations"
will be the tensor product of the two:
\beq
\begin{array}{ll}
\label{replgn}
\Pi: \widehat{{\cal L}}({\frak g}^{(n)})& \longrightarrow \rm{End}(V\otimes{\Bbb C}^{n+1})\\
\Pi(X\otimes p(\lambda))&\mapsto \Gamma(X)\otimes\rho(p(\lambda))
\end{array}
\eeq
Since this is the main object of the present work 
let us show in what follows how this construction generalises what has been already 
done for the affine Lie algebras \cite{K}, \cite{K1} and 
\cite{FLM}. In particular we are going to extend to our case what in the cited 
literature are called homogeneous basic representations, because in our opinion 
in this
 setting the construction of the generalised vertex operators turns out to be 
 more transparent. 
\par
More precisely let $Q$ be the root lattice associated with the
simple Lie algebra $\frak g$,  which we suppose to be of rank $l$ 
and let  $\Bbb C (Q)$ be its group algebra, i.e., the algebra with basis 
$e^\alpha$ $\alpha \in Q$ and multiplication:
$$
e^\alpha e^\beta=e^{\alpha+\beta}, \quad e^0=1.
$$ 
We shall denote by $\frak h=Q\otimes_{\Bbb Z}{\Bbb C}$ the complexification of $Q$ and by
$$
\widehat{\frak h}=\frak h \otimes \Bbb C (t,t^{-1})\oplus\Bbb C
$$
the affinization of $\frak h$, and finally by  $S$ the symmetric algebra over the space 
${\frak h}^{< 0}=\sum_{j< 0}\frak h\otimes t^j$ (following the literature we shall write 
$Ht^j$ in place of $H\otimes t^j$). Then we can define a representation $\pi$ of 
$\widehat{\frak h}$ on  $V_Q=S\otimes {\Bbb C}(Q)$  by setting $\pi=\pi_1\otimes \pi_2$ where $\pi_1$ acts 
on $S$ as 
\beq
\label{p1as}
\begin{array}{ll}
\pi_1(c_0)&=I\\ 
\pi_1(Ht^n)(At^s)&=\left\{ \begin{array}{ll} &HA t^{n-s} \mbox{ if $ n<0$}\\
& n\delta_{n,s}\langle H\vert A\rangle_{\frak g} \mbox{ if $ n\geq 0$}\end{array}\right.
\end{array}
\eeq
while $\pi_2$ act on $\Bbb (Q)$ simply by
\beq
\label{p2aq}
\pi_2(K)=0, \qquad \pi_2(Ht^n)e^\alpha =\delta_{n,0}\langle\alpha\vert H\rangle_{\frak g}e^\alpha. 
\eeq
Let now set $H_\alpha=\alpha\otimes1$,$\alpha(n)=\pi(H_\alpha t^n)$,  $H_n=\pi(Ht^n)$, and $e^\alpha$ 
the operator on $V_Q$ of multiplication 
by $1\otimes e^\alpha$ and consider the following $\mbox{End}(V_Q)$--valued fields:
\beq
\label{volfg}
\begin{array}{ll}
H(z)=&\sum_{n\in \Bbb Z}H_nz^{-n-1}\\
\Gamma_\alpha(z)=&\sum_{k=0}^n(\exp\left(\sum_{n\geq 1} \frac{\alpha(-n)z^n}{n}\right)
(\exp\left(\sum_{n\geq 1} \frac{\alpha(n)z^{-n}}{n}\right)e^\alpha z^\alpha
\end{array}
\eeq
Using these notations we can prove the 
\begin{theorem}\label{vorfgna} Let $V^n_Q=\bigoplus_{i=0}^n V_Q$ be the direct sum of 
$n+1$ copies of $V_Q$ then the following $\mbox{End}(V_Q^N)$--valued fields:
\beq
\label{vofgn}
\begin{array}{ll}
c_k&=c\Lambda^k\quad k=0,\dots,n \quad c\in {\Bbb C}\\
H^k_{\alpha}(z)&=\sum_{m \in \Bbb Z} \alpha(m)z^{-m-1}\Lambda^k=
H_\alpha(z) \Lambda^k
\quad k=0,\dots,n\\
\Gamma^k_\alpha(z)&=\sum_{k=0}^n(\exp\left(\sum_{m\geq 1} \frac{\alpha(-m)z^m}{m}\right)
(\exp\left(\sum_{m\geq 1} \frac{\alpha(m)z^{-m}}{m}\right)e^\alpha z^\alpha)\Lambda^k\\
&=\Gamma_\alpha(z)\Lambda^k \quad k=0,\dots,n
\end{array}
\eeq
define a vertex operator representation of the Lie algebra
$\widehat{{\cal L}}({\frak g})\otimes{\Bbb C}^{(n)}(\lambda)$, where the matrix $\Lambda$ is given by equation
\rref{efol}.
\end{theorem}
{\bf Proof } As usual we need only to check that our generating series satisfies 
the right OPE. But this can be easily done, keeping in mind the OPE of the fields 
\rref{volfg} (see \cite{K} and \cite{K1}).  We have indeed:
\beq
\label{opehhhy}
\begin{array}{ll}
H_\alpha^k(z)H^j_\beta(w)=&H_\alpha(z) H_\beta(w)\Lambda^k\Lambda^j=\\
&  \left\{\begin{array}{ll}
& \sim \frac{\langle H_\alpha\vert H_\beta\rangle}{(z-w)^2}\Lambda^{k+j}\mbox{ if $k+j\leq n$}\\
& 0 \mbox{ otherwise}\end{array}\right.\\
H^k(z)\Gamma^j_\alpha(w)=&H(z)\Gamma_\alpha(z)\Lambda^k\Lambda^j=\\
&  \left\{\begin{array}{ll} 
&\sim \frac{\langle H, \alpha\rangle}{z-w}\Gamma_{\alpha(z)}\Lambda^{k+j}\mbox{ if $k+j\leq n$}\\
& 0 \mbox{ otherwise}\end{array}\right.
\end{array}
\eeq
In similar way 
\beq
\label{opeyy}
\begin{array}{lll}
\Gamma_\alpha^k(z)\Gamma^j_\beta(w)=&\Gamma_\alpha(z)\Gamma_\beta(w)\Lambda^k\Lambda^j=0  
\hspace{1truecm}\mbox{ if $\alpha+\beta\notin  \Delta$}&\\
\Gamma_\alpha^k(z)\Gamma^j_\beta(w)=&\Gamma_\alpha(z)\Gamma_\beta(w)\Lambda^k\Lambda^j=&  \\
  & \left\{\begin{array}{ll} 
&\sim \epsilon (\alpha,\beta)\frac{\Gamma_{\alpha+\beta}}{z-w}\Lambda^{k+j}\mbox{ if $k+j\leq n$}\\
& 0 \mbox{ otherwise}\end{array}\right.  \mbox{  if $\alpha+\beta\in\Delta$}&\\ 
\Gamma_\alpha^k(z)\Gamma^j_{-\alpha} (w)=&\Gamma_\alpha(z)\Gamma_{-\alpha}(z)\Lambda^k\Lambda^j=& \\
 & \left\{\begin{array}{ll} 
&\sim \epsilon (\alpha,-\alpha)\frac{c_{k+j}}{(z-w)^2}+
\frac{\alpha(w)}{z-w}\Lambda^{k+j}
\hspace{.5truecm}\mbox{ if $k+j\leq n$}\\
& 0 \mbox{ otherwise}\end{array}\right.& 
\end{array}
\eeq 
 where with $\epsilon:Q\to \{\pm\}$ is  as  usual 2--cocycle (\cite{K1}) such that
$$
 \epsilon (\alpha,\beta) \epsilon (\beta,\alpha)=(-1)^{(\alpha\vert \beta)+(\alpha\vert \alpha)(\beta\vert \beta)}.
$$ 
\endpf    
\subsection{ Generalised boson--fermion correspondence}
In the next section we shall apply the theory of Kac Wakimoto \cite{KW}
on order to obtain a class of coupled soliton equations. Altought this theory may
be implemented  using directly the vertex operators given in Theorem 
\ref{vorfgna} even in this case there exists a generalised fermionic construction 
which is in our opinion worth to be presented at least in the case in which the 
simple Lie algebra $\frak g$ is of type $A$.\par 
Let us first  indeed define  the following (generalised polynomials) Clifford algebra
$\mbox{CL}^{(n)}$. 
\begin{defi}\label{dclpnd} Let $V^{n+1}$ denoted the infinite dimensional complex 
 vector space 
generated by the elements  $\{\psi^{(k)}_j,\psi^{*(k)}_j\}_{j\in \Bbb Z, n=0,\dots,n}$:
$$
V^{n+1}=\sum_{k=0}^n\left( \sum_{i\in \Bbb Z}{\Bbb C}\psi^{(k)}_i+\sum_{i\in \Bbb Z}{\Bbb C}\psi^{*(k)}_i \right)
$$
and let consider  on it the   symmetric bilinear form $\langle\cdot,\cdot\rangle_{V^{n+1}}$ defined 
by the relations
\beq
\label{blfonV}
\begin{array}{ll}
\langle\psi^{(k)}_l,\psi^{(j)}_m\rangle_{V^{n+1}}&=\langle\psi^{*(k)}_l,\psi^{*(j)}_m\rangle_{V^{n+1}}=0
\qquad k,j=0,\dots,n\quad k,j\in \Bbb Z\\
\langle\psi^{(k)}_l,\psi^{*(j)}_m\rangle_{V^{n+1}}&=
\left\{
\begin{array}{ll} \delta_{lm}  & k+j\leq n\\
0 & k+j>n.
\end{array}\right.
\end{array}
\eeq
Then  we call  polynomial Clifford algebra of length $n$ $\mbox{CL}^{(n)}$ the 
algebra given by
\beq
\label{clpnd}
\mbox{CL}^{(n)}=T(V^{n+1})/J^{n+1}
\eeq
where $T(V^{n+1})$ denotes the tensor algebra over $V^{n+1}$ and 
$J^{n+1}$ is the ideal 
generated by the elements $x\otimes y+y \otimes x-\langle x,y\rangle_{V^{n+1}}$ $x,y \in V$.
\end{defi}
Thus it is  easily checked that our definition implies that $\mbox{CL}^{(n)}$ 
is the algebra
generated as complex (infinite) vector space by the elements 
$\{\psi^{(k)}_j,\psi^{*(k)}_j\}_{j\in \Bbb Z, n=0,\dots,n}$ which satisfy the following relations:
\beq
\label{clalr}
\begin{array}{ll}
&\psi^{(k)}_i\psi^{(j)}_l+\psi^{(j)}_l\psi^{(k)}_i=0\qquad \psi^{*(k)}_i\psi^{*(j)}_l+\psi^{*(j)}_l\psi^{*(k)}_i=0\\
&\psi^{(k)}_i\psi^{*(j)}_l+\psi^{*(j)}_l\psi^{(k)}_i=\left\{ \begin{array}{ll}
&\delta_{il}\Lambda^{j+k}\mbox{ if $j+k\leq n$}\\
& 0 \mbox{ otherwise.}\end{array}\right.
\end{array}
\eeq
The algebra $\mbox{CL}^{(n)}$ possesses a representation on  the direct sum of $n+1$ 
copies of infinite wedge algebras: 
\beq
\label{iwsnp1}
F^{(n)}=\oplus_{i=0}^nF^i
\eeq
where the spaces $F^i$ with $i=0,\dots,n$ are copies of the infinite wedge space $F$
generated by the semi--infinite monomials
$$
\underline{i}_1\land \underline{i}_2\land\dots\land\underline{i}_j\land\dots
$$
where the $i_j$ are  integers such that 
$$
i_1>i_2>i_3\qquad \mbox{and } i_j=i_{j-1}-1 \quad \mbox{for $n$ big enough}
$$
(see \cite{K} for more details). Every space $F^i$ has a charge decomposition 
$$
F^i=\bigoplus_{m\in \Bbb Z}F^i_m
$$
by letting 
$$
\vert m\rangle^i=(\underline{m} \land \underline{m-1}\land \underline{m-2}\land\dots)^i
$$
denote the vacuum vector of charge $m$ in $F^i$ and $F^i_m$ the linear space 
spanned by all semi--infinite 
monomials in $F^i$, which differ from $\vert m\rangle^i$ only at finite number of places, in the same 
way we can decompose $F^{(n)}$ as $F^{(n)}=\bigoplus_{m\in \Bbb Z}F^{(n)}_m$ where $F^{(n)}_m=
\oplus_{i=0}^nF^i_m$.
On $F^{(n)}$ we define respectively the 
following action
of $\psi^{(k)}_i$ and $\psi^{*(k)}_i$:
\beq
\begin{array}{ll}
\psi^{(k)}_i((\underline{i}_1\land&\underline{i}_2\land\dots)^0,\dots,
(\underline{i}_1\land\underline{i}_2\land\dots)^j,\dots
 (\underline{i}_1\land \underline{i}_2\land\dots)^n)\\
&=\sum_{l=0}^{n-k}\psi_ie_{l+k,l}
((\underline{i}_1 \land \underline{i}_2\dots)^0,\dots,
(\underline{i}_1\land\underline{i}_2\land\dots)^j,\dots
  (\underline{i}_1\land\underline{i}_2\land\dots)^n)\\
&=(\underbrace{0,\dots,0}_{k},(\psi_i(\underline{i}_1\land\underline{i}_2\land\dots)^0)^k,
\dots,(\psi_i(\underline{i}_1\land\underline{i}_2\land\dots)^j)^{j+k},\dots\\
&\phantom{(\underbrace{0,\dots,0}_{k},(\psi_i(\underline{i}_1\land\underline{i}_2\land
\dots)^0)^k,\dots}{\dots,(\psi_i(\underline{i}_1\land \underline{i}_2\land\dots)^{n-k})^n} \\
\psi^{*(k)}_i((\underline{i}_1\land&\underline{i}_2\land\dots)^0,\dots,
(\underline{i}_1\land\underline{i}_2\land\dots)^j,\dots
 (\underline{i}_1\land \underline{i}_2\land\dots)^n)\\
&=\sum_{l=0}^{n-k}\psi^*_ie_{l+k,l}
((\underline{i}_1 \land \underline{i}_2\dots)^0,\dots,
(\underline{i}_1\land\underline{i}_2\land\dots)^j,\dots
  (\underline{i}_1\land\underline{i}_2\land\dots)^n)\\
&=(0,\dots,0,(\psi^*_i(\underline{i}_1\land\underline{i}_2\land\dots)^0)^k,\dots
(\psi^*_i(\underline{i}_1\land\underline{i}_2\land\dots)^j)^{j+k},\dots\\
&\phantom{(0,\dots,0,(\psi^*_i(\underline{i}_1\land\underline{i}_2\land\dots)^0)^k,\dots}
\dots,(\psi^*_i(\underline{i}_1\land \underline{i}_2\land\dots)^{n-k})^n) \\
\end{array}
\eeq 
where the action of the operators $\psi_j$ and $\psi^*_j$ is given by the usual formula
\cite{K}:
\beq
\label{psiact}
\begin{array}{ll}
\psi_j((\underline{i}_1\land \underline{i}_2\land\dots)^{m})=\left\{ \begin{array}{ll}
 0 \mbox{ if $j= i_s $ for some $s$}&\\
 (-1)^s(\underline{i}_1\land\dots\land\underline{i}_s^m\land 
\underline{i}_{i_s+1}\land\dots)^{m} \mbox{ if $i_s>j>i_{s+1}$}&\end{array}\right.&\\
\psi_j^*((\underline{i}_1\land \underline{i}_2\land\dots)^m)=\left\{ \begin{array}{ll}
 0 \mbox{ if $j\neq i_s $ for all $s$}&\\
 (-1)^{s+1}(\underline{i}_1\land\dots\land\underline{i}_{s-1}\land 
\underline{i}_{i_s+1}\land\dots)^m \mbox{ if $j=i_s$}.&\end{array}\right.& 
\end{array}
\eeq
A simple computation shows indeed that the operators defined above 
satisfies equations \rref{clalr} giving arise to a representation of $CL^{(n)}$.
It is clear that this representation is indecomposable and that the vector 
$$
\vert 0\rangle=((0\land-1\land-2\land\dots)^0,\underbrace{0,\dots,0}_{n})
$$
is a cyclic vector for it (i.e., $F^{(n)}=CL^{(n)}(\vert 0\rangle)$)
which satisfies the relations
$$
\psi^{(k)}_j\vert 0\rangle=0 \mbox{ for $j\leq 0$,}\qquad
\psi^{*(k)}_j\vert 0\rangle=0  \mbox{ for $j> 0$}\qquad   k=0,\dots,n.
$$
Moreover it can be also checked that their action preserves the 
bilinear (non degenerated but not positive definite) form on $F^{(n)}$ given by 
\beq
\label{blfofn}
\begin{array}{ll}
\langle (\underline{i}_1\land \underline{i}_2\land\dots,)^{0},&\dots,
(\underline{i}_1 \land \underline{i}_2\land\dots,)^{n}\vert
(\underline{i}_1\land \underline{i}_2\land\dots)^{0},\dots,(\underline{i}_1
\land \underline{i}_2\land \dots)^n\rangle_{F^{(n)}}\\
&=\sum_{l=0}^n\sum_{m=0}^l\langle\underline{i}_1\land \underline{i}_2\land\dots)^l\vert
(\underline{i}_1\land \underline{i}_2\land\dots)^{l-m}\rangle
\end{array}
\eeq
where $\langle\cdot\vert\cdot\rangle$ denote the Hermitian form on $F$ for 
which the canonical basis for $F$ 
is orthonormal. We have indeed: 
$$
\begin{array}{ll}
&\langle (\psi^{(k)}_i(\underline{i}_1\land \underline{i}_2\land\dots)^0,\dots,(\underline{i}_1\land
\underline{i}_2\land\dots)^n)\vert
(\underline{i}_1\land\underline{i}_2\dots)^0,\dots,(\underline{i}_1\land
\underline{i}_2\land\dots)^n,)\rangle_{F^{(n)}}\\
&=\langle(0,\dots,0,(\psi_i(\underline{i}_1\land\underline{i}_2\land\dots)^0)^k,
\dots,(\psi_i(\underline{i}_1\land \underline{i}_2\land\dots)^{n-k})^n\vert \\
&\phantom{=\langle(0,\dots,0,(\psi_i(\underline{i}_1\land\underline{i}_
2\land\dots)^0)^k,\dots }
(\underline{i}_1\land\underline{i}_2\land\dots)^0,\dots,(\underline{i}_1\land
\underline{i}_2\land\dots)^n,)\rangle_{F^{(n)}}\\
&=\sum_{l=0}^{n-k}\sum_{m=0}^l
\langle\psi_i(\underline{i}_1\land \underline{i}_2\land\dots)^l\vert
(\underline{i}_1\land \underline{i}_2\land\dots)^{l-m}\rangle\\
&=\sum_{l=0}^{n-k}\sum_{m=0}^l
\langle(\underline{i}_1\land \underline{i}_2\land\dots)^l \vert
\psi^*_i(\underline{i}_1\land \underline{i}_2\land\dots)^{l-m})\rangle\\
&=\langle(\underline{i}_1\land \underline{i}_2\land\dots)^0,\dots,(\underline{i}_1\land
\underline{i}_2\land\dots)^n \vert \psi^{*(k)}_i((\underline{i}_1\land 
\underline{i}_2\land\dots)^0,
\dots,(\underline{i}_1\land\underline{i}_2\land\dots)^n)\rangle_{F^{(n)}}.
\end{array}
$$
The more significative consequence of the equations \rref{clalr} are the following 
commutation relations:
\beq
\label{cmrfpsn}
\begin{array}{ll}
\left[\psi^{(l)}_i\psi^{*(m-l)}_j,\psi^{(p)}_k\right]&=\left\{\begin{array}{ll} \delta_{kj}\psi^{(m+p)}_i 
\mbox{ if $m+p\leq n$}&\\  0 \mbox{ otherwise}&\end{array}\right.\\
\left[\psi^{(l)}_i\psi^{*(m-l)}_j,\psi^{*(p)}_k\right]&=\left\{\begin{array}{ll} -\delta_{ki}\psi^{*(m+p)}_j 
\mbox{ if $m+p\leq n$}&\\  0 \mbox{ otherwise,}&\end{array}\right.
\end{array}
\eeq
which can be checked as follows
$$
\begin{array}{ll}
\left[\psi^{(l)}_i\psi^{*(m-l)}_j,\psi^{(p)}_k\right]&=\psi^{(l)}_i\psi^{*(m-l)}_j\psi^{(p)}_k-\psi^{(p)}_k
\psi^{(l)}_i\psi^{*(m-l)}_j\\
&=
\psi^{(l)}_i\psi^{*(m-l)}_j\psi^{(p)}_k+\psi^{(l)}_i\psi^{(p)}_k\psi^{(m-l)*}_j-\psi^{(l)}_i\psi^{(p)}_k\psi^{(m-l)*}_j\\
&\phantom{=}-\psi^{(p)}_k\psi^{(l)}_i\psi^{*(m-l)}_j\\
&=\psi^{(l)}_i\delta_{jk}\sum_{s=0}^{n-s}e_{s+j+k,s}+\psi^{(p)}_i\psi^{(l)}_k\psi^{(m-l)*}_j
-\psi^{(p)}_k\psi^{(l)}_i\psi^{(m-l)*}_j\\
&\stackrel{\rref{clalr}}{=}\delta_{jk}\psi_i\sum_{r=0}^{n-r}e_{r+l,r}
\sum_{s=0}^{n-s}e_{s+j+k,s}\\
&=\left\{ \begin{array}{ll} \delta_{kj}\psi^{(m+p)}_i 
\mbox{ if $m+p\leq n$}&\\  0 \mbox{ otherwise}&\end{array}\right.
\end{array}
$$
while the similar proof for the second one  is left to the reader.\par
The importance of equations \rref{cmrfpsn} is due to the fact that can be used to 
define a representation of a ``polynomial'' generalisation ${\frak g\frak l}^{(n)}_\infty$
of the infinite dimensional Lie algebra ${\frak g\frak l}_\infty$ 
\begin{defi}\label{dglni} Let ${\frak g\frak l}^{(n)}_\infty$ be the infinite 
dimensional Lie algebra given by the tensor product
\beq
\label{glnitp}
{\frak g\frak l}^{(n)}_\infty={\frak g\frak l}_\infty\otimes{\Bbb C}^{(n)}(\lambda)
\eeq
where ${\Bbb C}^{(n)}(\lambda)$ is the nilpotent polynomial ring \rref{quotspc}.
\end{defi}
The reader will immediately recognise that this definition is nothing else that
the generalisation to the infinite dimensional case of the construction of polynomial
Lie algebras given in the second section. Therefore a moment's reflection
shows that, recalling the definition of the 
algebra ${\frak g\frak l}_\infty$,  we can reformulate the above definition saying that 
${\frak g\frak l}^{(n)}_\infty$ is the Lie algebra given by the linear span of the basis
$\{E^k_{ij}\}_{j,i\in \Bbb Z,k=0,\dots,n}$  with Lie brackets given by the formulas:
\beq
\label{lbfglin}
\left[E^k_{ij},E^s_{lm}\right]=\left\{\begin{array}{ll} \delta_{jl}E^{k+s}_{im}- 
\delta_{im}E^{k+s}_{lj}\mbox{ if $k+s\leq n$} &\\
0\mbox{ otherwise.}\end{array}\right.
\eeq
Similarly, starting from equations \rref{clalr}, one can view   the polynomial 
Clifford algebra $CL^{(n)}$ as the tensor product $CL\otimes{\Bbb C}^{(n)}(\lambda)$.\par
Being defined our infinite dimensional polynomial Lie algebra
${\frak g\frak l}^{(n)}_\infty$ we can construct its fermionic representation
on $F^{(n)}$.
\begin{theorem}\label{repglin} The map $\Psi:{\frak g\frak l}^{(n)}_\infty\to 
\rm{End}(F^{(n)})$
given by:
\beq
\label{gilnr}
\Psi(E_{ij}^k)=\frac{1}{k+1}\sum_{l=0}^k\psi^{(k-l)}_i\psi^{*(l)}_j\qquad i,j\in \Bbb Z\quad 
k=0,\dots,n
\eeq
defines a representation $\Psi$ of ${\frak g\frak l}^{(n)}_\infty$ on $F^{(n)}$.
\end{theorem}
{\bf Proof } 
Using formulas \rref{cmrfpsn} we have for $k+s\leq n$:
$$
\begin{array}{ll}
\left[\Psi(E_{ij}^k),\Psi(E_{lm}^s)\right]=&\frac{1}{k+1}\frac{1}{s+1}
\left[\sum_{p=0}^k\psi_i^{(k-p)}\psi_j^{*(p)},\sum_{q=0}^{s}\psi^{(s-q)}_l\psi^{*(q)}\right]\\
&=\frac{1}{k+1}\frac{1}{s+1}\sum_{p=0}^k\sum_{q=0}^s(\left[\psi_i^{(k-p)}\psi_j^{*(p)},
\psi^{(s-q)}_l\psi^{*(q)}\right])\\
&=\frac{1}{k+1}\frac{1}{s+1}\sum_{p=0}^k\sum_{q=0}^s(\left[\psi_i^{(k-p)}\psi_j^{*(p)},
\psi^{(s-q)}_l\right]\psi^{*(q)}_m)\\
&\phantom{=}+\frac{1}{k+1}\frac{1}{s+1}\sum_{p=0}^k\sum_{q=0}^s(\psi_i^{(k-p)}\left[\psi_j^{*(p)},
\psi^{(s-q)}_l\psi^{*(q)}\right])\\
&=\frac{1}{s+1}\frac{1}{k+1}\sum_{p=0}^k\sum_{q=0}^s(\delta_{jl}
\psi_i^{(k+s-q)}\psi^{*(q)}_m-\delta_{mi}\psi_l^{(s-q)}\psi^{*(k+q)}_j)\\
&=\delta_{jl}\frac{1}{s+1}\sum_{q=0}^s\psi_i^{(k+s-q)}\psi^{*(q)}_m-
\delta_{mi}\frac{1}{s+1}\sum_{q=0}^k \psi_l^{(s-q)}\psi^{*(k+q)}_j)\\
&=\delta_{jl}\Psi(E^{k+s}_{im})-\delta_{mi}\Psi(E^{k+s}_{lj})=\Psi(\left[E^{k}_{ij},E^s_{lm}\right])
\end{array}
$$
since it is easily checked that 
$\frac{1}{s+1}\sum_{q=0}^k \psi_l^{(s-q)}\psi^{*(k+q)}_j=\Psi(E^{k+s}_{lj})$ and\\ 
$\frac{1}{s+1}\sum_{q=0}^s\psi_i^{(k+s-q)}\psi^{*(q)}_m=\Psi(E^{k+s}_{im})$.
While if $k+s>n$ a similar computation gives 
$\left[\Psi(E_{ij}^k),\Psi(E_{lm}^s)\right]=0$ as wanted.
\endpf
Note that, while the action of $\rm{CL}^{(n)}$ interchanges the charges,
every subspaces $F^{(n)}_m$ is left invariant by 
 the representation $\Psi$. Further these latter spaces are indecomposable so that $\Psi$ is the
 direct
sum of its restrictions on $F^{(n)}_m$. 
\par
We are actually also interested in the corresponding group representation, despite the fact 
that 
${\frak g}{\frak l}^{(n)}_\infty=\rm{Lie}(GL_\infty\ltimes{\frak g}{\frak l}_\infty)$ the exponential 
map of this algebra lies  in a bigger group, which contains $(GL_\infty\ltimes{\frak g}{\frak l}_\infty)$
as a proper subgroup \cite{PS}. 
\begin{prop}\label{lgfglni} Let $N^{(n)}_\infty$ be the following subset of 
${\frak g}{\frak l}_\infty^{(n)}$:
$$
N^{(n)}_\infty=\{ I+X\vert\ X\in 
{\frak g}{\frak l}_\infty\otimes\lambda{\Bbb C}^{(n-1)}(\lambda)\}
$$
then 
\begin{enumerate}
\item $N^{(n)}_{\infty}$ is a group.
\item The group $G^{(n)}_\infty=GL_\infty\ltimes N^{(n)}_\infty$ is the littlest group which contains the
image of the exponential map on ${\frak g}{\frak l}^{(n)}_\infty$.
\end{enumerate}
\end{prop}
{\bf Proof } 
\begin{enumerate}
\item It is enough to show that for every element $Z=I+X\in N^{(n)}$
there exists (actually unique) element $Y\in {\frak n}^{(n)}=
{\frak g}{\frak l}_\infty\otimes\lambda{\Bbb C}^{(n-1)}
(\lambda)$ such that \\$Z=\exp(Y)$. But since this algebra  is also a
associative ring we can compute for every $X$ in it the expression   
$Y=\ln(1+X)=\sum_{k=1}^n (-1)^{k+1}\frac{X^k}{k}$ 
where the sum turn out to be finite because $X^{m}=0$ if $m > N$. Then a direct computation
 shows 
that $I+X=\exp(Y)$.
\item The second statement follows immediately once one recognises that  $GL_\infty$ is the 
exponential 
group of the Lie algebra ${\frak g}{\frak l}_\infty$ (\cite{K}) and that 
${\frak g}{\frak l}_\infty^{(n)}={\frak g}{\frak l}_\infty\ltimes{\frak n}^{(n)}$.
\end{enumerate}
\endpf
The representation $\Psi$ can be exponentiated to the Lie group
$G^{(n)}_\infty$. Namely if $g$ is an element of $G^{(n)}_\infty$ of the form
$g=\exp{X_0}$ where $X_0$ belongs to ${\frak g}{\frak l}_\infty\otimes 1\simeq 
{\frak g}{\frak l}_\infty$ 
then we have the natural extension of the usual case \cite{K}
\beq
\label{grpl0}
\begin{array}{ll}
\Psi(g)&((\underline{i}_1\land \underline{i}_2\land\dots)^0,\dots,
(\underline{i}_1\land\underline{i}_2\land \dots)^j,\dots
  (\underline{i}_1\land\underline{i}_2\land\dots)^n)\\
&=((g(\underline{i}_1\land \underline{i}_2\land\dots)^0)^0,\dots,
(g(\underline{i}_1\land\underline{i}_2\land \dots)^j)^j,\dots
  (g(\underline{i}_1\land\underline{i}_2\land\dots)^n)^n)
\end{array}
\eeq
where 
$$
(g(\underline{i}_1\land \underline{i}_2\dots)^s)^s=
(\sum_{j_1>j_2>\dots}\det(g^{i_1,i_2,\dots}_{j_1,j_2,\dots})(j_1\land j_2\dots))^s.
$$
While if $g$ is an element of $G^{(n)}_\infty$ of the form
$g=\exp{X_k}$ with $X_k$ which belongs to ${\frak g}{\frak l}_\infty \otimes \lambda^k$ with $k>0$
then the action becomes: 
\beq
\label{glnigdr}
\begin{array}{ll}  
\Psi(\exp{X_k})&((\underline{i}_1\land \underline{i}_2\dots)^0,\dots,
(\underline{i}_1\land\underline{i}_2\land \dots)^j,\dots
  (\underline{i}_1\land\underline{i}_2\land\dots)^n)\\
&=(\underline{i}_1\land\underline{i}_2\land\dots)^0,\dots,
(\underline{i}_1\land\underline{i}_2\land\dots)^{k-1},\\
&(\underline{i}_1\land \underline{i}_2\land \dots)^k+X_k(\underline{i}_1\land 
\underline{i}_2\land \dots)^0)^k,\dots,\\
& \dots,(\sum_{q=0}^{[\frac{r}{k}]}\frac{1}{q!}X^q_k
(\underline{i}_1\land\underline{i}_2\land\dots)^{r-kq})^r,\dots,\\
  &\phantom{\dots}\dots,((\sum_{q=0}^{[\frac{n}{k}]}\frac{1}{q!}X^q_k
\underline{i}_1\land\underline{i}_2\land\dots)^{n-kq})^n
\end{array}
\eeq
where we have denoted with $[\frac{r}{k}]$ the integer part of $\frac{r}{k}$.\par
It is still possible to
construct  a bosonization of the representation $\Psi$ of ${\frak g}{\frak l}^{(n)}_\infty$, 
which generalise to the our contest that already known in 
the literature \cite{K}, \cite{K1},\cite{KW}.
To achieve this task we have first to extend the representation $\Psi$ from 
${\frak g}{\frak l}^{(n)}_\infty$ to 
${\frak a}^{(n)}_\infty={\frak a}_\infty\otimes{\Bbb C}^{(n)}$,
this requires to get rid from anomalies to modify our representation $\Psi$ by putting
\beq
\label{mrpsfai}
\Psi(E^k_{ij})=\left\{ \begin{array}{ll} \frac{1}{k+1}\sum_{l=0}^k \psi^{(k-l)}_i\psi^{*(l)}_j
\mbox{ if $i\neq j$ or $i=j>0$}&\\
-\frac{1}{k+1}\sum_{l=0}^k\psi^{*(l)}_j\psi^{(k-l)}_i \mbox{ if $i=j\leq 0$.}&\end{array}\right. 
\eeq
Next we must define the subalgebra ${\frak s}^{(n)}$ spanned by the elements 
$$
s^{k}_i=\sum_{j\in \Bbb Z}E^{k}_{j,j+i},\quad \mbox{ and } c^k\qquad k=0,\dots,n 
$$
whose Lie brackets are
\beq
\label{heislb}
\left[s^{k}_p,s^{j}_q\right]=\left\{ \begin{array}{ll} p\delta_{p,-q}c^{k+j}
\mbox{ if $j+k\leq n$}&\\
0\mbox{ otherwise.}&\end{array}\right. 
\eeq
Using the representation $\Psi$ this algebra is given by the 
free bosonic fields $\alpha^k_j$:
\beq
\label{fbfaf}
\begin{array}{lll}
\alpha^k_j&=\frac{1}{k+1}\sum_{l=0}^k\sum_{i\in \Bbb Z}\psi^{(k-l)}_i\psi^{*(l)}_{i+j}
\phantom{\sum_{l=0}^k\sum_{-i\in \Bbb N}} \mbox{if $j\in \Bbb Z \{0\}$},&k=0,\dots,n\\
\alpha^k_0&=\frac{1}{k+1}\sum_{l=0}^k\sum_{i>0}\psi^{(k-l)}_i\psi^{*(l)}_{i}
-\sum_{l=0}^k\sum_{i\leq 0}\psi^{(k-l)}_i\psi^{*(l)}_{i}&\quad k=0,\dots,n.
\end{array}
\eeq
We have indeed:
\beq
\label{hbla}
 \left[\alpha^k_r,\alpha^j_s\right]=\left\{\begin{array}{ll} r\delta_{r,-s}\Lambda^{k+j}
\mbox{ if $k+j\leq n$}&\\
0 \mbox{ otherwise.} &\end{array}\right.
\eeq
Now following Kac in \cite{K} we introduce the bosonic Fock space $B^{(n)}$
given by the direct of $n+1$ copies of the usual  bosonic fock space 
$B=\Bbb C[x_1,x_2,\dots;q,q^{-1}]$:
\beq
\label{bfspn}
B^{(n)}=\bigoplus_{i=0}^n B_i
\eeq
where $B_i$ $i=0,\dots,n$ are copies of $B$. 
For our purposes it is also useful to look at this space as the tensor product
between the Fock space $B$ and an $n+1$ dimensional complex space:
\beq
\label{bfspntp}
B^{(n)}=B\otimes{\Bbb C}^{(n+1)}
\eeq 
in what follows we shall also need the decomposition in ``charged subspaces''
of $B^{(n)}$ given by:
\beq
\label{dbnicsm}
B^{(n)}=\bigoplus_{m\in \Bbb Z}B_m\qquad B_m=q^m{\Bbb C}[x_1,x_2,\dots]\otimes{\Bbb C}^m.
\eeq
In this setting it is namely easy to define a representation 
$r^{B^{(n)}}$ of the Heisenberg algebra $\frak s^{(n)}$ on $B^{(n)}$ as:
\beq
\label{rohbab}
\begin{array}{ll}
\left.
\begin{array}{ll}
&r^{B^{(n)}}(s_m^k)=
\frac{\partial}{\partial x_m}\otimes \Lambda^k\\
&r^{B^{(n)}}(s_{-m}^k)=mx_m\otimes\Lambda^k 
\end{array}
\right\}
&\mbox{ if $m>0$} \quad
k=0,\dots,n\\
\quad r^{B^{(n)}}(s_0^k)=
q\frac{\partial}{\partial q}\otimes\Lambda^k& k=0,\dots,n.
\end{array}
\eeq
It is straightforward to see that the  usual isomorphism of 
$\frak s={\frak s}^{(0)}$--modules $\sigma :F\simeq B$ (see \cite{K}) can be extended
to a ${\frak s}^{(n)}$--modules $\sigma^n:F^{(n)}\simeq B^{(n)}$  simply by taking
the direct sum of $n+1$ copies of the isomorphism $\sigma$.\par
Let us now introduce in the contest of the fermionic fields the generating
series:
\beq 
\label{gsffnf}
\psi^{(k)}(z)=\sum_{j\in \Bbb Z}\psi^{(k)}_jz^j\qquad \psi^{*(k)}(z)=\sum_{j\in \Bbb Z}\psi^{*(k)}_jz^{-j}
\quad k=0,\dots,n
\eeq
and also the following operator
\beq
\label{gsfbgn}
\Gamma_+^k(z)=\sum_{n\geq 1}\alpha^k_n\frac{z^{-n}}{n}\qquad  
\Gamma_-^k(z)=\sum_{n\geq 1}\alpha^k_{-n}\frac{z^n}{n}
\quad k=0,\dots,n.
\eeq
In the bosonic picture using the tensor product these latter generating 
series can be also written as:
\beq
\label{gsibtp}
\Gamma_+^k(z)=\sum_{n\geq 1}\frac{z^{-n}}{n}\frac{\partial}{\partial x_n}\otimes\Lambda^k \qquad  
\Gamma_-^k(z)=\sum_{n\geq 1}\frac{z^n}{n}x_n\otimes\Lambda^k.
\eeq
The isomorphism $\sigma^n:F^{(n)}\simeq B^{(n)}$ allows us to construct a generalised
boson--fermionic correspondence:
\begin{theorem}\label{gbfcnt} For every $k=0,\dots,n$ we have:
\beq
\label{fbfcit}
\begin{array}{ll}
\psi^{(k)}(z)&=z^{\alpha^k_0}q\Gamma_-^k(z)\Gamma_+^k(z)^{-1}\\
\psi^{*(k)}(z)&=q^{-1}z^{-\alpha^k_0}\Gamma_-^k(z)^{-1}\Gamma_+^k(z).
\end{array}
\eeq
\end{theorem}
{\bf Proof }
Let us prove only the first of equations \rref{fbfcit} since a completely similar 
construction works for the second ones.
From the equations \rref{fbfaf} and \rref{gsffnf} we obtain that
$$
\begin{array}{ll}
\left[\alpha^k_j,\psi^{(i)}(z)\right]&=\left\{ \begin{array}{ll} 
z^j\psi^{(k+i)}(z) \mbox{ if $i+k\leq n$}&\\
 0 \mbox{ otherwise}&\end{array}\right.\\
\left[\alpha^k_j,\psi^{*(i)}(z)\right]&=\left\{ \begin{array}{ll} -z^j\psi^{*(k+i)}(z) 
\mbox{ if $i+k\leq n$}&\\
0 \mbox{ otherwise}&\end{array}\right.
\end{array}
$$
now using the map $\sigma^{(n)}$ we can transport these relations to $B^{(n)}$ we have 
indeed for $j>0$ we have
$$
\begin{array}{ll}
\sigma^{(n)}\left[\alpha^k_j,\psi^{(i)}(z)\right](\sigma^{(n)})^{-1}&=
\left[\frac{\partial}{\partial x_j}\otimes\Lambda^k,\sigma^{(n)}\psi^{(i)}(z)(\sigma^{(n)})^{-1}
\right]\\
&=\left\{ \begin{array}{ll} z^j\sigma^{(n)}\psi^{(k+i)}(z)(\sigma^{(n)})^{-1}
 \mbox{ if $i+k\leq n$}&\\
0 \mbox{ otherwise}\end{array}\right.
\end{array}
$$
while for $j$ negative 
$$
\begin{array}{ll}
\sigma^{(n)}\left[\alpha^k_j,\psi^{(i)}(z)\right](\sigma^{(n)})^{-1}&=
\left[x_j\otimes \Lambda^k,\sigma^{(n)}\psi^{(i)}(z)(\sigma^{(n)})^{-1}
\right]\\
&=\left\{ \begin{array}{ll} \frac{z^{-j}}{j}\sigma^{(n)}\psi^{(k+i)}(z)(\sigma^{(n)})^{-1}
 \mbox{ if $i+k\leq n$}&\\
0 \mbox{ otherwise.}\end{array}\right.
\end{array}
$$
Using these  equations,  the fact that $\psi^{(k)}_j$ can be written as 
$\psi_j\Lambda^k$ and lemma 14.5  of \cite{K} we can now conclude that
the operator $\sigma^{(n)}\psi_j^{(k)}(\sigma^{(n)})^{-1}$ brings the subspace $B^{(n)}_m$ 
in the subspace $B^{(n)}_{m+1}$ for every $m$ and it  is of the form 
$$
\sigma^{(n)}\psi_j^{(k)}(\sigma^{(n)})^{-1}=C_m(z)q\Gamma^k(z)
$$
with
$$
\Gamma^k(z)=\left\{\exp\left(\sum_{j\leq 1}z^jx_j\right)\exp\left(-\sum_{j\leq 1}\frac{z^{-j}}{j}
\frac{\partial}{\partial x_j}\right)\right\}\otimes\Lambda^k
$$
while the same  argument used in the proof of Theorem 14.10 in \cite{K} shows that
$C_m(z)=z^{m+1}$. 
\endpf
\begin{theorem}\label{rebain} The generating series for the representation 
$\Psi$ \rref{gilnr} of ${\frak g}{\frak l}^{(n)}_\infty$  is
\beq
\label{gsfpsin}
\sum_{i,j\in \Bbb Z}z_1^iz_2^{-j}\Psi(E^k_{ij})=\left(\frac{z_1}{z_2}\right)^m\frac{1}
{1-\frac{z_2}{z_1}}\Gamma^k(z_1,z_2)
\eeq
where
\beq
\label{gkz1z2} 
\Gamma^k(z_1,z_2)=
(k+1)\exp(\sum_{p\geq 1}
(z^p_1-z^p_2)x_p)\exp(\sum_{p\geq 1}\frac{z^{-p}_1-z^{-p}_2}{p}\frac{\partial}{\partial x_p})\Lambda^k.
\eeq
\end{theorem} 
{\bf Proof } We observe that from \rref{gsffnf} follows 
$$
\sum_{i,j\in \Bbb Z}z_1^iz_2^{-j}\Psi(E^k_{ij})=\frac{1}{k+1}\sum_{l=0}^k\psi_i^{(k-l)}(z_1)
\psi_j^{*(l)}(z_2)
$$
substituting \rref{fbfcit} we get
$$
\sum_{i,j\in \Bbb Z}z_1^iz_2^{-j}\Psi(E^k_{ij})=\frac{1}{k+1}\sum_{l=0}^k
z_1^{\alpha^{k-l}_0}q\Gamma_-^{k-l}(z_1)\Gamma_+^{k-l}(z_1)^{-1}
q^{-1}z_2^{-\alpha^l_0}\Gamma_-^l(z_2)^{-1}\Gamma_+^l(z_2).
$$
Since it holds  \cite{K} for every $0\leq l\leq k$ that  
$$
\Gamma_+^{k-l}(z_1)^{-1}\Gamma_-^l(z_2)^{-1}=\Gamma_-^l(z_2)^{-1}\Gamma_+^{k-l}(z_1)^{-1}
\left(1-\frac{z_2}{z_1}\right)^{-1}
$$
the previous equation using also \rref{gsibtp} becomes
$$
\begin{array}{ll}
\left(1-\frac{z_2}{z_1}\right)^{-1}
\frac{1}{k+1}\sum_{l=0}^k
\sum_{s=0}^{n-k+l}z_1^me_{k-l+s,s}\sum_{r=0}^{n-l}z_2^{-m}e_{l+s,s}&\\
\sum_{s=0}^{n-k+l}\exp(\sum_{p\geq 1}(z^p_1x_p)e_{k-l+s,s}
\sum_{r=0}^{n-l}\exp(\sum_{p\geq 1}-z^p_2x_p)e_{l+r,r}&\\
\sum_{s=0}^{n-k+l}\exp(\sum_{p\geq 1}\frac{z^{-p}_1\frac{\partial}{\partial x_p}}{p})e_{k-l+s,s}
\sum_{r=0}^{n-l}\exp(\sum_{p\geq 1}\frac{-z^{-p}_1\frac{\partial}{\partial x_p}}{p}) e_{l+r,r}&
\end{array}
$$
and finally
$$
(k+1)\left(\frac{z_1}{z_2}\right)^m\left(1-\frac{z_2}{z_1}\right)^{-1}
\exp(\sum_{p\geq 1}
(z^p_1-z^p_2)x_p)\exp(\sum_{p\geq 1}\frac{z^{-p}_1-z^{-p}_2}{p}\frac{\partial}{\partial x_p})\Lambda^k
$$
\endpf
\section{Coupled Hirota bilinear equations}
The aim of this section is to derive from the vertex operator algebras constructed
in the previous one the corresponding hierarchies of Hirota bilinear equations.
The key link to connect our representations  with the corresponding  bilinear equations
are opportune  homogeneous Casimir operators acting on particular tensor product 
of representations.
Since the Lie algebras ${\frak g}^{(n)}$ and the generalised Clifford
algebra $\rm{CL}^{(n)}$ as well have an ad--invariant symmetric bilinear non degenerate
form we can use them in order to define  a corresponding homogeneous Casimir operator.
\begin{defi} 
\begin{enumerate}
\item Let $X_i$ $i=1,\dots, \dim(\frak g)$ be a basis for $\frak g$ we define   
\beq
\begin{array}{ll}
\Omega&=\sum_{k,h =0}^n\sum_{i,j=1}^{\dim(\frak g)}\sum_{p,q\in \Bbb Z}\frac{1}{h+k+1}
\langle X^k_i\otimes t^p,X^h_j\otimes t^q\rangle(X^k_i\otimes t^p)\otimes (X^k_i\otimes t^q)\\
&=\sum_{k=0}^n\sum_{l=0}^k\sum_{i,j=1}^{\dim(\frak g)} \\
&\phantom{\sum}\left[\sum_{p,q\in \Bbb Z}\frac{1}{k+1}
\langle X^{k-l}_i\otimes t^p,X^l_j\otimes t^q\rangle(X^{k-l}_i\otimes t^p)\otimes (X^l_i\otimes t^q)\right]
\end{array}
\label{csopgn}  
\eeq
\item 
\beq
\label{casopfp}
\begin{array}{ll}
\Omega_1&=\sum_{m=0}^n\sum_{k=0}^n\sum_{i,j\in \Bbb Z}\frac{1}{m+k+1}
\langle \psi^{(m)}_i,\psi^{*(k)}_j\rangle_{V^{n+1}}\psi^{(m)}_i\otimes\psi^{*(k)}_j\\
&=\sum_{k=0}^n\sum_{l=0}^k\sum_{j\in\Bbb Z}\frac{1}{k+1}\psi_j^{(k-l)}\otimes\psi_j^{*(l)}.
\end{array}
\eeq
\end{enumerate}
\end{defi}
Observe that these Casimir operators can be defined as the canonical Casimir operators 
of the respectively reduced tensor product
\beq
\label{rtpcln}
\begin{array}{ll}
(\widehat{{\frak g}}\otimes {\Bbb C}^{(n)}(\lambda))\otimes_{{\Bbb C}^{(1)}(\lambda)}
(\widehat{{\frak g}}\otimes {\Bbb C}^{(n)}(\lambda))&\\
(\rm{CL}\otimes {\Bbb C}^{(n)}(\lambda))\otimes_{{\Bbb C}^{(1)}(\lambda)}(\rm{CL}\otimes {\Bbb C}^{(n)}(\lambda))&
\end{array}
\eeq
since for example we have 
$$
(\sum_{k=0}^n\sum_{j\in \Bbb Z}\psi_j\otimes \lambda^k)\otimes_{\Bbb C^{(1)}(\lambda)}(\sum_{k=0}^n\sum_{j\in \Bbb Z}\psi^*_j\otimes \lambda^k)
=(\sum_{k=0}^n\sum_{j\in \Bbb Z}(\psi_j\otimes\psi^*_j)\otimes \lambda^k=\Omega_1.
$$
This in turn suggests  to consider the action of the Casimir operators on a similar 
modified tensor product of representation's space. 
 The first step towards the construction of such modified tensor product
 is to note (as already observed 
in  equation \rref{ltegcn} for the algebras ${\frak g}^{(n)}$) that the representation's
spaces can be viewed as the tensor product between a infinite dimensional space $V$ 
and ${\Bbb C}^{n+1}$ where the polynomial ring ${\Bbb C}^{(n)}(\lambda)$ acts.
Moreover it is easily to see that this ${\Bbb C}^{(n)}(\lambda)$--module is isomorph to
${\Bbb C}^{(n)}(\lambda)$  thought as  ${\Bbb C}^{(n)}(\lambda)$--module over itself.
This fact allows us to decompose our representation's space as the tensor product:
\beq
\label{tplr}
V^{(n)}(\lambda)=V\otimes {\Bbb C}^{(n)}(\lambda).
\eeq
Then mimicking  what done  in \rref{rtpcln} we consider the ``modified'' tensor product
\beq
\label{mtpgn}
(V\otimes {\Bbb C}^{(n)}(\lambda))\otimes_{{\Bbb C}^{(1)}}(V\otimes {\Bbb C}^{(n)}(\lambda))
\eeq
where the representation of our algebra still survive, because this process boils down 
to perform a projection with respect to an invariant subspace namely:
$$
\sum_{i,j=1 \atop{i+j>n}}^n(V\otimes{\Bbb C}^{(i)}(\lambda))\otimes((V\otimes{\Bbb C}^{(j)}(\lambda)).
$$
These will be the space where our generalised Hirota equations will live.
\subsection{Coupled KP hierarchies}
Let us compute these equations  explicitly starting with the case of 
${\frak g}{\frak l}^{(n)}_\infty$.
\par
Using equations \rref{cmrfpsn} it is possible to check that $\Omega_1$ commute 
with action of  ${\frak g}{\frak l}^{(n)}_\infty$ on $F^{(n)}(\lambda)\otimes_{{\Bbb C}^{(1)}} 
F^{(n)}(\lambda)$ and therefore with
that of $GL^{(n)}_\infty$. But this in turn says that any element $\tau=(\tau_0,\dots,\tau_n)$ of 
the orbit of 
 $GL^{(n)}_\infty(\vert 0\rangle,0,\dots,0)$ satisfies the equation
\beq
\label{ombieqifs}
\sum_{p,q=0\atop{p+q\leq n}}^n\sum_{k=0}^n\sum_{l=0}^k\sum_{j\in\Bbb Z}
\frac{1}{k+1}\psi_j^{(k-l)}(\tau_p)\otimes\psi_j^{*(l)}(\tau_q)=0
\eeq
and to the contrary  using the argument of  Theorem 14.11 in \cite{K} that holds
\begin{lem}\label{glinforb} The orbit of $GL^{(n)}_\infty\vert 0\rangle$ is the set of 
all nonzero solutions $\tau\in F^{(n)}_0$ of equation \rref{ombieqifs}.
\end{lem}
Our generalised Hirota bilinear equation will be the bosonic version of equation 
\rref{ombieqifs}. To apply to it the isomorphism $\sigma^{(n)}$ we have to write it 
in terms of $\psi^{(k)}(z)$ and  $\psi^{*(k)}(z)$ as
\beq
\label{hbefoz}
z^0\mbox{--term of }\sum_{p,q=0\atop{p+q\leq n}}^n
 \sum_{k=0}^n\sum_{l=0}^k\frac{1}{k+1}\psi^{(k-l)}(z)\tau_p\otimes\psi^{*(l)}(z)\tau_q=0.
\eeq
Then its bosonizated form is
\beq
\label{hglibn}
\res_{z=0}\sum_{p,q=0\atop{p+q\leq n}}^n\sum_{k=0}^n\sum_{l=0}^k(\exp\sum_{j\geq 1}z^j(x'_{j}-x_{j}''))
(\exp-\sum_{j\geq 1}\frac{z^{-j}}{j}(\frac{\partial}{\partial x'_{j}}-\frac{\partial}{\partial x_{j}''}))\tau_p(x')\tau_q(x'')=0.
\eeq
Introducing the new variables 
$$
x_j=\frac12(x'_j+x''_j)\qquad y_j=\frac12(x'_j-x''_j)
$$
equation \rref{hglibn} becomes:
$$
\res_{z=0}\sum_{p,q=0\atop{p+q\leq n}}^n\sum_{k=0}^n\sum_{l=0}^k(\exp 2\sum_{j\geq 1}z^j(y_j))
(\exp-\sum_{j\geq 1}\frac{z^{-j}}{j}(\frac{\partial}{\partial y_{j}}))\tau_p(x+y)\tau_q(x-y)=0.
$$
This latter equation can be easily written in terms of elementary Schur polynomials
as:
\beq
\label{hblsp}
\sum_{p=0}^k\sum_{j\geq 0} S_j(2y)S_{j+1}(-\tilde{\partial}_y)\tau_p(x+y)\tau_{k-p}(x-y)=0\qquad k=0,\dots,n
\eeq
where as usual $\tilde{\partial}_y$ means $\left(\frac{\partial}{\partial y_1},\frac12 \frac{\partial}{\partial y_2},\frac13
\frac{\partial}{\partial y_3},\dots\right)$.
Then introducing the Hirota bilinear differentiation by:
$$
P(D_1,D_2,\dots)fg=P(\frac{\partial}{\partial u_1},\frac{\partial}{\partial u_2},\dots)
f(x_1+u_1,x_2+u_2,\dots)g(x_1-u_1,x_2-u_2,\dots)
$$
and using the Taylor formula
$$
\begin{array}{ll}
P(\tilde{\partial}_y)\tau_p(x+y)\tau_q(x-y)&=P(\tilde{\partial}_u)\tau_p(x+y+u)\tau_q(x-y-u)\vert_{u=0}\\
&=P(\tilde{\partial}_u)\left(\exp\sum_{j\geq 1}y_j\frac{\partial}{\partial u_j}\right)\tau_p(x+u)\tau_q(x-u)\vert_{u=0}.
\end{array}
$$
we can write \rref{hblsp} in the Hirota bilinear form: 
\beq
\label{hbeff}
\sum_{p=0}^k\sum_{j\geq 0}S_j(2y)S_{j+1}(-\tilde{D})
\left(\exp\sum_{s\geq 1}y_sD_s\right)\tau_p\tau_{k-p}\quad k=0,\dots,n.
\eeq
(Here again as usual $\tilde{D}$ stands for $(D_1,\frac12 D_2,\frac13 D_3,\dots0)$.
Expanding \rref{hbeff} as a multiple Taylor series in the variables $y_1,y_2,\dots$
we obtain that each coefficient of the series must vanish giving arise to a hierarchy
infinite number of non linear partial differential equation in a Hirota bilinear form,
which of course contains the celebrated KP hierarchy. 
Observe that   
$P(D_1,\dots,D_k)\sum_{p=0}^k\tau_p\tau_{k-p}=0$ identically for any odd
monomial $P(D_1,\dots,D_k)$ in the Hirota operators $D_k$  
because $\sum_{p=0}^k\tau_p\tau_{k-p}=\sum_{p=0}^k\tau_{k-p}\tau_p$ for any $k=0,\dots,n$.
Therefore the first non trivial coupled Hirota equations are:
\beq
\label{fbhegc}
\begin{array}{ll}
&(D^4_1+3D^2_2-4D_1D_3)\tau_0\tau_0=0\\
&(D^4_1+3D^2_2-4D_1D_3)\tau_0\tau_1=0\\
&\dots=\dots\\
&(D^4_1+3D^2_2-4D_1D_3)(\sum_{p=0}^k\tau_p\tau_{k-p}))=0\\
&\dots=\dots\\
&(D^4_1+3D^2_2-4D_1D_3)(\sum_{p=0}^n\tau_p\tau_{n-p}))=0
\end{array}
\eeq
To write this equations in the ``soliton variables''  we  perform the change of variables 
$u_0=2\frac{\partial^2 \log(\tau_0)}{\partial x^2}$,
$u_i=\frac{\tau_i}{\tau_0}$ which generalises to our case those proposed 
by Hirota, Hu and Tang   in \cite{HXT}. In these new variables equations \rref{fbhegc} read
\beq
\label{HHTen}
\left \{ \begin{array}{ll}
&\frac34u_{0yy}-(u_{0t}-\frac32u_0u_{0x}-\frac14u_{0xxx})_x=0\\
&\\
& u_{kxxxx}-4u_{kxt}+3u_{kyy}+6u_0u_{kxx} \\ 
& +\left(\sum_{j=1}^{k-1} 2u_{jx}u_{(k-j)t}+2u_{jt}u_{(k-j)x}-3u_{jy}u_{(k-j)y}\right.\\
&\phantom{+\sum_{j=1}^k}-6u_{0}u_{jx}u_{(k-j)x} -2u_{jxxx}u_{(k-j)x}\\
&\left.\phantom{+\sum_{j=1}^k} -3u_{jxx}u_{(k-j)xx}-2u_{jx}u_{(k-j)xxx}\right)=0 \quad k=1,\dots n
 \end{array} \right.
\eeq 
where $x=x_1$, $y=x_2$ and $t=x_3$.
The vertex operator construction offers a canonical way to produce a class of generalised
soliton solutions for these equations. Let $u^j_1, \dots u^j_N$ $v^j_1, \dots v^j_N$
$j=0,\dots,n$ be some 
indeterminates, then using  the property of the Taylor expansions we have that
written in component
\beq
\label{gggkpn}
\ba{ll}
&(\prod_{j=N}^1(\sum_{k_j=0}^n\Gamma(u^{k_j}_j,v^{k_j}_j)\Lambda^{k_j})
(\tau_0(x_1,x_2,\dots),\dots,\tau_n(x_1,x_2,\dots))^T)_m\\
&=\sum_{k_1,\dots,k_N,s=0 \atop{k_1+\dots+k_N+s=m}}^n
 \prod_{1\leq i\leq j\leq N}\left[ \frac{(u_j^{(k_j)}-u_i^{(k_i)})
(v_j^{(k_j)}-v_i^{(k_i)})}{(u_j^{(k_j)}-v_i^{(k_i)})(v_j^{(k_j)}-u_i^{(k_i)})} \right.\\
&\quad \times \left(\exp \sum_{r\geq1}\sum_{l=1}^N((u_l^{k_l})^r-(v_l^{k_l})^r)x_r\right)\\
&\left. \phantom{\frac{1}{2}}\quad \times \tau_s(\dots,x_r-\frac1r\sum_{l=1}^N
 ((u_l^{k_l})^{-r}-(v_l^{k_l})^{-r}),\dots)\right] \qquad  m=0,\dots,n.
\ea
\eeq
This equation tells us that any matrix $\sum_{k=0}^n\Gamma(u^k,v^k)\Lambda^k$ acts as nilpotent
operator it hold indeed 
\begin{lem}\label{nlpingn} For every $s$ $0\leq s\leq n$ we have that
\beq
\label{mniee}
(\sum_{k=0}^n\Gamma(u^k,v^k)\Lambda^k)^s(0,\dots,\tau_r,0,\dots)^T=0
\eeq
for every $\tau_r\in {\Bbb C}(x_1,x_2,\dots)$ and every choice of $u^k$ and $v^k$,
  if and only if $s>\left[\frac{1+\sqrt{1+8(n-r)}}{2} \right]$ where $[x]$ denotes 
the integer part of $x$. Moreover the m--th component of the vector 
$(\sum_{k=0}^n\Gamma(u^k,v^k)\Lambda^k)^s(0,\dots,\tau_r,0,\dots)^T$ vanishes identically 
if and only if $s>\left[\frac{1+\sqrt{1+8(m-r)}}{2} \right]$.
\end{lem}
{\bf Proof } If we set in formula \rref{gggkpn}  $u_j^{k_j}=u^{k_j}$ and  
$v_j^{k_j}=v^{k_j}$
then it is easily to check that $(\sum_{k=0}^n\Gamma(u^k,v^k)\Lambda^k)^s(0,\dots,\tau_r,0,\dots)^T=0$
for every $\tau_s\in {\Bbb C}(x_1,x_2,\dots)$ and every choice of $u^{k_j}$ and $v^{k_j}$
if and only if $k_j=k_i$ for some $k_j$ and $k_i$ in each set of non 
positive integers $\{k_1,\dots k_s\}$ which appears in the right hand of 
\rref{gggkpn}. Or in other words if and only if
any set of non negative integers  $\{k_1,\dots,k_s\}$ such that 
$\sum_{i=1}^sk_i=n-r$ contains at least two elements which coincide.
Suppose that $\{k_1,\dots,k_s\}$ is a sequence with all elements distinct such that
$\sum_{i=1}^sk_i=n-r$, then,  since the sequence of $s$ non negative pairwise distinct 
integers whose sum is the smallest is obviously $\{0,1,\dots,N\}$,
we must have that $n-r=\sum_{i=1}^sk_i\geq\sum_{i=0}^si=\frac{s(s-1)}{2}$. Therefore
equation   \rref{mniee} is identically satisfied only and only if 
$s>\left[\frac{1+\sqrt{1+8(n-r)}}{2} \right]$.
A completely similar argument proves the second part of the lemma.
\endpf
Using the statement of the lemma we can write the exponential map of an element 
of the type $\sum_{k=0}\alpha_k\Gamma(u^k,v^k)$ as 
$$
\exp(\sum_{k=0}\alpha_k\Gamma(u^k,v^k))=\sum_{k=0}^n\Lambda^k
\left(\sum_{j=0}^{\left[\frac{1+\sqrt{1+8k}}{2}\right]}
\frac{1}{j!}\sum_{s_1+\dots+s_j=k}\prod_{i=1}^j\alpha_{s_i}\Gamma(u_{s_i},v_{s_i})\right).
$$ 
Therefore from the lemma \ref{glinforb} follows that the $N$ soliton solution of the
polynomial KP hierarchy is
\beq
\label{nsolsolckp}
\begin{array}{ll}
&\tau_{\alpha^1_0,\dots,\alpha^1_n,\dots, \alpha^N_n,u^1_0,\dots,u^N_n,v^1_0,\dots,v^N_n}(x)\\
&=\sum_{k=0}^n\Lambda^k\left(\sum_{j=0}^{\left[\frac{1+\sqrt{1+8k}}{2}\right]}
\frac{1}{j!}\sum_{s_1+\dots+s_j=k}\prod_{i=1}^j\alpha_{s_i}\Gamma(u_{s_i},v_{s_i})\right)(1,0,\dots,0)^T.
\end{array}
\eeq
In particular the 1--soliton solution (again written in component) is
$$
\begin{array}{ll}
&(\tau_{\alpha_0,\dots,\alpha_n,u_0,\dots,u_n,v_0,\dots,v_n}(x))_m\\
&=\sum_{j=0}^{\left[\frac{1+\sqrt{1+8k}}{2}\right]}
\frac{1}{j!}\sum_{s_1+\dots+s_j=k}\prod_{i=1}^j\alpha_{s_i}\prod_{0\leq i<l\leq j}\frac{(u_{s_i}-u_{i_l})
(v_{s_i}-v_{s_l})}{(u_{s_i}-v_{s_l})( v_{s_i}-u_{s_l})}\\
&\times \left(\exp \sum_{r\geq1}\sum_{i=1}^j(u_{s_i}^r-v_{s_{i}}^r)x_r\right)\qquad m=0,\dots,n.
\end{array}
$$
In the simplest coupled case when $n=1$ equations \rref{HHTen} become 
\beq
\label{HHTe}
\begin{array}{ll}
\frac34u_{0yy}-(u_{0t}-\frac32u_0u_{0x}-\frac14u_{0xxx})_x&=0\\
 u_{1xxxx}-4u_{1xt}+3u_{1yy}+6u_0u_{1xx}&=0.
\end{array}
\eeq
For this equations the 1--soliton solution with  
$\alpha_0=\alpha_1=1$ takes the form
$$
\left( \begin{array}{c}  \tau_0\\ \tau_1\end{array}\right)=
\left( \begin{array}{c} 1+\exp(\sum_{r\geq 1}(u_0^r-v_0^r)x_r)\\ 
\begin{array}{cc}&\exp(\sum_{r\geq 1}(u_1^r-v_1^r)x_r)\\
&+2\frac{(u_0-u_1)(v_0-v_1)}{(u_0-v_1)(v_0-u_1)}
\exp(\sum_{r\geq 1}(u_0^r-v_0^r+u_1^r-v_0^r)x_r)\end{array}
\end{array}\right).
$$
Of course in the contest of the single equations \rref{HHTe} (and actually for $n=2$)
we can view the indeterminates $x_4,x_5\dots$ as parameters in the expression of the
solution, which will explicitly depend only from the first three ones. Therefore explicitly:
\beq
\label{1sosoln2}
\begin{array}{ll}
u_0(x,y,t)=&\frac12(u_0-v_0)(\cosh(\frac12(u_0-v_0)x+(u_0^2-v_0^2)y+(u_0^3-v_0^3)t
+\gamma_0))^{-2}\\
u_1(x,y,t)=&\frac12\left(\cosh(\frac12(u_0-v_0)x+(u_0^2-v_0^2)y+(u_0^3-v_0^3)t
+\gamma_0)\right)^{-1}\\
&\times\{ e^{-\frac12((u_0-v_0)x+(u_0^2-v_0^2)y+(u_0^3-v_0^3)t+\gamma_0}\\
&+2\frac{(u_0-u_1)(v_0-v_1)}{(u_0-v_1)(v_0-u_1)}
e^{\frac12((u_0-v_0)x+(u_0^2-v_0^2)y+(u_0^3-v_0^3)t+\gamma_0)}\} \\
&\times e^{\frac12((u_1-v_1)x+(u_1^2-v_1^2)y+(u_1^3-v_1^3)t+\gamma_1)}
\end{array}
\eeq
where $\gamma_i$ with $i=0,1$ are arbitrary constants.
\subsection{Coupled KdV and Boussinesq hierarchies}
Similarly we may construct a generalisation of the KdV hierarchy (i.e., a coupled KdV 
hierarchy) by considering the principal ``basic'' representation of the polynomial Lie algebra
$\widehat{{\cal L}}({\frak s}{\frak l}_2^{(n)})$.  From what done in section 4 we consider 
the $\hat{{\cal L}}({\frak s}{\frak l}_2^{(n)})$--module $V^n_Q=\oplus_{j=0}^n{\Bbb C}(
x_1,x_3,x_5,\dots)$ given by the formulas
\beq
\label{brsl2n}
\begin{array}{llll}
H^k_j&=\Lambda^k\frac{\partial}{\partial x_j},\quad   H_{-j}^k&=jx_j\Lambda^k\quad & j\in {\Bbb N}^{\rm{odd}},
k=0,1\dots,n\\
c_k&=\Lambda^k, \quad 2d-\frac12A_0^k&=-\sum_{j\in {\Bbb N}^{\rm{odd}}}jx_j\Lambda^k\frac{\partial}{\partial x_j} &
k=0,1\dots,n\\
A^k(z)&=\frac12(\Gamma^k(z)-1), & & k=0,1\dots,n\\
\end{array}
\eeq
where 
$$
H^k_{2j+1}=t^j\left(X^k_\alpha-tX^k_{-\alpha}\right),\quad 
A^k_{2j}= -t^j\left(tH^k_{\alpha}\right),\quad 
A^k_{2j+1}= t^j\left(X^k_\alpha-tX^k_{-\alpha}\right)
$$
with
$$
X^k_\alpha=\left(\begin{array}{cc} 0 & 1\\ 0 & 0\end{array}\right)\otimes\lambda^k\quad
X^k_{-\alpha}=\left(\begin{array}{cc} 0 & 0\\ 1 & 0\end{array}\right)\otimes\lambda^k\quad
H^k_\alpha=\left(\begin{array}{cc} 1 & 0\\ -1 & 0\end{array}\right)\otimes\lambda^k
$$
and finally
$$
\Gamma^k(z)=\left(\exp 2\sum_{j\in {\Bbb N}^{\rm{odd}}}z^jx_j\right)
\left(\exp -2\sum_{j\in {\Bbb N}^{\rm{odd}}}\frac{z^{-j}}{j}\frac{\partial}{\partial x_j}\right)
\Lambda^k.
$$
Then the polynomial Hirota bilinear equation are given by
$$
\Omega(v\otimes_{{\Bbb C}^{(1)}(\lambda)}v)=\mu v\otimes_{{\Bbb C}^{(1)}(\lambda)}v,
$$
where $\mu\in {\Bbb C}$ which is equivalent to the following hierarchy of bilinear equations:
\beq
\label{hblkdv}
\begin{array}{lc}
&\sum_{p=0}^k\left( \sum_{j>0}S_j(4y_1,0,4y_3,\dots)S_j(-\frac21 D_1,0,-\frac23 D_3,\dots)-
8\sum_{j\in {\Bbb N}^{\rm{odd}}}jy_jD_j \right)\\
&\times\left(\exp\sum_{j\in {\Bbb N}^{\rm{odd}}}y_jD_j\right) \tau_p\tau_{k-p}=0\quad k=0,\dots,n.
\end{array}
\eeq 
Reasoning as in the previous case of the coupled KP equations we have that the first
non trivial bilinear equations in the hierarchy are: 
$$
\sum_{p=0}^k(-4D_1D_3+D_1^4)\tau_p\tau_{k-p}=0\quad k=0,\dots,n\\
$$
are a  generalisation of the coupled equation (12) in \cite{HXT} namely by imposing
the variable transformation $u_0=2\frac{\partial^2 \log (\tau_0)}{\partial x^2}$ ,
$u_i=\frac{\tau_i}{\tau_0}$ $i=1,\dots,n$ we get:
\beq
\label{cKdVn}
\left \{ \begin{array}{ll}
&(u_{0t}-\frac32u_0u_{0x}-\frac14u_{0xxx})_x=0\\
&\\
& u_{kxxxx}-4u_{kxt}+6u_0u_{kxx} \\ 
& +\left(\sum_{j=1}^{k-1} 2u_{jx}u_{(k-j)t}+2u_{jt}u_{(k-j)x}\right.\\
&\phantom{+\sum_{j=1}^k}-6u_{0}u_{jx}u_{(k-j)x} -2u_{jxxx}u_{(k-j)x}\\
&\left.\phantom{+\sum_{j=1}^k} -3u_{jxx}u_{(k-j)xx}-2u_{jx}u_{(k-j)xxx}\right)=0 \quad k=1,\dots n.
 \end{array} \right.
\eeq 
In particular for $n=1$ we have
\beq
\label{cKdV2}
\begin{array}{ll}
u_{0t}-\frac32u_0u_{0x}-\frac14u_{0xxx}=0&\\
6u_0u_{1xx}+u_{1xxxx}-4u_{1xt}=0.&
\end{array}
\eeq
This latter equations make the contact with the literature \cite{HXT} and 
\cite{Sak} (more precisely  setting $u=u_0$, 
$v=u_{1x}$ and rescaling the time $t \to -4t$ one obtains  equations (2) of \cite{Sak}).
Further is worth to note that by taking the derivative with respect to $x$ of the 
second equation and putting $v_0=u_0$ and $v_1=u_{1xx}$ equations \rref{cKdV2}
become
$$
\begin{array}{ll}
v_{0t}=\frac32 v_0v_{0x}+\frac14 v_{0xxx}&\\
v_{1t}=\frac14 v_{1xxx}+\frac32 v_0v_{1x}+\frac32 v_{0x}v_{1}.&
\end{array}
$$
These equations  are bihamiltonian with respect the two Poisson tensors \cite{AF3} \cite{FRTS}
$$
\begin{array}{ll}
P_1&=\left(\begin{array}{cc} \frac12 \partial_{xxx}+2v_0\partial_x+v_{0x} & 0\\
0& -2\partial_x\end{array}\right) \\
P_2&=\left(\begin{array}{cc} 0 & \frac12 \partial_{xxx}+2v_0\partial_x+v_{0x} \\
\frac12 \partial_{xxx}+2v_0\partial_x+v_{0x} -2\partial_x& 2v_1\partial_x+v_{1x}\end{array}\right)
\end{array}
$$
namely
$$
\left(\begin{array}{c} v_{0t}\\v_{1t}\end{array}\right)=P_1
\left(\begin{array}{c} -\frac12 v_0\\-\frac18 v_{1xx}-\frac34 v_0v_1 \end{array}\right)=
P_2\left(\begin{array}{c} \frac12 v_1\\-\frac12 v_0 \end{array}\right).
$$
Similarly they can be also written in the Lax form $\frac{dL}{dt}=[L,P]$ where 
$$
\begin{array}{ll}
L&=\left(\begin{array}{cc}\partial_{xx}+v_0 & 0\\
v_1 & \partial_{xx}+v_0\end{array}\right)\\
P&=\left(\begin{array}{cc}-\partial_{xxx}-\frac34v_{0x}-\frac32 v_0\partial_x& 0\\
-\frac34v_{1x}-\frac32 v_1\partial_x &-\partial_{xxx}-\frac34v_{0x}-\frac32 v_0\partial_x \end{array}\right).
\end{array}
$$
Moreover analogous changes of variables lead to the Lax pairs for the other hierarchies
arising from the Lie algebras $\widehat{\cal L}({\frak s}{\frak l}_k^{(n)})$.
\par
Actually, as in the standard case, these hierarchy can be recovered from the polynomial
KP \rref{fbhegc} by performing a reduction procedure, which amounts to eliminate the 
dependence from the ``even'' variables $x_{2j}$ $j\in {\Bbb N}$ of the Fock space. 
This in turn corresponds 
to restrict the representation of ${\frak g}{\frak l}^{(n)}_\infty$ onto its subalgebras
$\hat{{\cal L}}({\frak s}{\frak l}_2^{(n)})$, giving Lie algebraic explanation   
of what done in the recent literature  \cite{SK}.
Therefore the soliton solution for the coupled KdV hierarchies can be recovered 
from those written for the coupled KP equations \rref{nsolsolckp} erasing the even variables.
In the particular case whereas $n=2$ this reduction method
applied to \rref{1sosoln2} leads to the following solutions:
\beq
\label{KdV1sosoln2}
\begin{array}{ll}
u_0(x,y,t)=&\frac12(u_0-v_0)(\cosh(\frac12(u_0-v_0)x+(u_0^3-v_0^3)t
+\gamma_0))^{-2}\\
u_1(x,y,t)=&\frac12\left(\cosh(\frac12(u_0-v_0)x+(u_0^3-v_0^3)t
+\gamma_0)\right)^{-1}\\
&\times\{ e^{-\frac12((u_0-v_0)x+(u_0^3-v_0^3)t+\gamma_0}\\
&+2\frac{(u_0-u_1)(v_0-v_1)}{(u_0-v_1)(v_0-u_1)}
e^{\frac12((u_0-v_0)x+(u_0^3-v_0^3)t+\gamma_0)}\} \\
&\times e^{\frac12((u_1-v_1)x+(u_1^3-v_1^3)t+\gamma_1)}
\end{array}
\eeq
where $\gamma_i$ with $i=0,1$ are still arbitrary constants.
\par
In the same way we can recovered from the coupled KP hierarchy the ``coupled Boussinesq''
hierarchy by erasing all the variables $x_{3j}$ with $j\in {\Bbb N}$, which again corresponds
to restrict our representation to the Lie algebra 
$\widehat{{\cal L}}({\frak s}{\frak l}_3^{(n)})$.
In this case the first non trivial bilinear Hirota equations:
$$
(D^4_1+3D^2_2)(\sum_{p=0}^n\tau_p\tau_{n-p})=0\qquad k=0,\dots,n.
$$
In particular when $n=1$ putting $u_0=2(\log(\tau_0))_{xx}$ and as usual $u_1=\frac{\tau_1}{\tau_0}$ 
we get
$$
\begin{array}{ll}
&3u_{0tt}+u_{0xxxx}+6u^2_{0x}+6u_0u_{0xx}=0\\
&3u_{1tt}+u_{1xxxx}+6u_0u_{1xx}=0
\end{array}
$$
where we have put $x=x_1$ $t=x_2$. The multi--soliton solutions of these equation can 
obviously recovered from the solutions \rref{nsolsolckp} by erasing the variables $x_{3j}$.
\subsection{Coupled BKP hierarchies and their first reductions}
The construction presented above can be extended to simple  Lie algebras, which are not
of type $A$. In particular we would like to finish the paper by outlining briefly the case
of the Lie algebras of type $B$. 
In order to construct the bilinear Hirota equations for the polynomial BKP hierarchy,
we have to consider the polynomial Clifford algebra $\rm{CL}_B^{(n)}$ defined (like 
in definition \ref{dclpnd}) as the 
Clifford algebra generated by the elements $\phi_i^{(k)}$ $i\in \Bbb Z$ $k=0,\dots,n$ 
and by the symmetric bilinear form:
$$
\langle \phi^{(k)}_i,\phi^{(j)}_l\rangle=\left\{ \begin{array}{ll}
&(-1)^i\delta_{i,-l}\mbox{ if $j+k\leq n$}\\
& 0 \mbox{ otherwise.}\end{array}\right.
$$
Let $V$ be the irreducible Verma module with highest weight vector $\vert 0\rangle$ 
for the usual Clifford Lie algebra $\rm{CL}_B$
(i.e. $\rm{CL}_B^{(0)}$). The algebra $\rm{CL}_B^{(n)}$ acts on $V^{(n)}=\oplus_{k=0}^nV_k$,
$V_k\simeq V$ for all $k$ by the formula
$$
\phi^{(j)}_i(v_0,\dots,v_n)=(\underbrace{0,\dots,0}_{j},\phi_iv_0,\dots,\phi_iv_{n-j}).
$$
Using this action we can define for $n \in {\Bbb Z}^{\rm{odd}}$ the neutral bosonic fields:
$$
\beta^k_m=\frac{1}{2(k+1)}\sum_{l=0}^k\sum_{j\geq 1}(-1)^{j+1}\phi^{(k-l)}_j\phi^{(l)}_{-j-m}
$$
which generate the associated Heisenberg algebra 
$$
\left[\beta^h_p,\beta^k_q\right]=\left\{ \begin{array}{ll}
&\frac12p\delta_{p,-q}\Lambda^{h+k}\mbox{ if $h+k\leq n$}\\
& 0 \mbox{ otherwise.}\end{array}\right.
$$
Now we can define a generalised boson--fermion correspondence of type $B$ 
$\sigma^{(n)}_B:V^{(n)}\to  B^{(n)}=\oplus_{k=0}^nB_k$ where 
$B_k={\Bbb C}[x_1,x_3,x_5,\dots;q]/(q^2-\frac12)$ for all $k$, which nothing else that the 
direct sum of $n+1$ copies of the usual isomorphism \cite{K} $\sigma_B$ and therefore
$$
\begin{array}{c}
\sigma_B^{(n)}(\underbrace{0,\dots,0}_{k-1},\vert 0\rangle ,0\dots,0)=
(\underbrace{0,\dots,0}_{k-1},1 ,0\dots,0)\\
\sigma_B^{(n)}\phi^k_0(\vert 0\rangle ,0\dots,0)=
(\underbrace{0,\dots,0}_{k-1},q ,0\dots,0)
\end{array}
$$ 
and for $p\in {\Bbb N}^{\rm{odd}}$
$$
\sigma_B^{(n)}\beta^k_p(\sigma_B^{(n)})^{-1}=\Lambda^k\frac{\partial}{\partial x_p}\qquad 
\sigma_B^{(n)}\beta^k_{-p}(\sigma_B^{(n)})^{-1}=\frac12\Lambda^k px_p.
$$
Then  if we introduce the neutral fermionic fields:
$$
\phi^{(k)}(z)=\sum_{i\in \Bbb Z}\phi_i^{(k)}z^i
$$
we can show (as in the case of ${\frak a}^{(n)}_\infty$) that
$$
\sigma_B^{(n)}\phi^{(k)}(z)(\sigma_B^{(n)})^{-1}=\Lambda^kq\exp\left(\sum_{j\in {\Bbb N}^{\rm{odd}}}x_jz^j\right)
\exp\left(-2\sum_{j\in {\Bbb N}^{\rm{odd}}}\frac{z^{-j}}{j}\frac{\partial}{\partial x_j}\right).
$$
Our aim is now to construct a fermionic representation of the infinite dimensional 
polynomial Lie algebra ${\frak s}{\frak o}^{(n)}_\infty={\frak s}{\frak o}_\infty\otimes{\Bbb C}^{(n)}(\lambda)$
(and actually of ${\frak b}^{(n)}_\infty={\frak b}_\infty\otimes{\Bbb C}^{(n)}(\lambda)$)
spanned by the elements $F^k_{ij}=(-1)^jE^k_{ij}-(-1)^iE^k_{-j,-i}$ where the $E^k_{ij}$
are the basis of ${\frak g}{\frak l}^{(n)}_\infty$ previously considered.
Mimicking  the same proof of Theorem \ref{repglin} one  can prove indeed that the following 
formula:
$$
\rho(F^k_{ij})=\frac{1}{k+1}\sum_{l=0}^k\phi^{(k-l)}_i\phi^{(l)}_{-j}
$$
defines a representation of ${\frak s}{\frak o}^{(n)}_\infty$, which can be linearly extended
to a representation of ${\frak b}^{(n)}_\infty$ by putting
$$
\begin{array}{ll}
\hat{\rho}(F^k_{ij})&=\left\{\begin{array}{ll}\frac{1}{k+1}\sum_{l=0}^k\phi^{(k-l)}_i\phi^{(l)}_{-j}
\mbox{ if $i\neq j$ or $i=j>0$}&\\
\frac{1}{k+1}\sum_{l=0}^k\phi^{(k-l)}_i\phi^{(l)}_{-j}-\frac12\Lambda^k
\mbox{ if $i= j<0$ }&\end{array}\right.\\
\hat{\rho}(c_k)&=\Lambda^k \qquad k=0,\dots,n.
\end{array}
$$
This representation turns out 
to be the direct sum of two representation defined respectively on $V^{(n)}_0$
(the even elements $V^{(n)}$) and on $V^{(n)}_1$ (the odd ones). Moreover it can be 
checked that the map  $\sigma_B^{(n)}:V^{(n)}_0\simeq \oplus_{k=0}^n B_{k0}$ (where 
$B_{k0}={\Bbb C}(x_1,x_2,x_3,\dots)$ for all $k$) is a 
${\frak s}{\frak o}^{(n)}_\infty$--isomorphism between the representation $\rho_{\vert V_0}$ and the 
following vertex operator construction of the same  algebra:
$$
\sum_{i,j\in \Bbb Z}z^i_1z^{-j}_2F^k_{ij}\mapsto\frac12\; \frac{1-z_2/z_2}{1+z_2/z_1}(\Gamma_B^k(z_1,z_2)-1)
$$
where 
$$
\Gamma_B^k(z_1,z_2)=(k+1)\Lambda^k\exp\left(\sum_{j\in {\Bbb N}^{\rm{odd}}}x_j(z_1^j+z_2^j)\right)
\exp\left(-2\sum_{j\in {\Bbb N}^{\rm{odd}}}\frac{z_1^{-j}-z_2^{-j}}{j}\frac{\partial}{\partial x_j}\right).
$$
In order to construct the polynomial nBKP hierarchy of Hirota bilinear equation 
we use the operator
$$
\Omega_1^B=\sum_{k=0}^n\frac{1}{k+1}\sum_{l=0}^k\sum_{j\in \Bbb Z}(-1)^j\phi^{(k-l)}_j\otimes\phi^{(l)}_{-j}
$$
commuting with the action of the algebra ${\frak b}^{(n)}_\infty$.
The equation on $V^{(n)}\otimes_{{\Bbb C}^{(1)}(\lambda)}V^{(n)}$
$$
\Omega_1^B(\tau\otimes_{{\Bbb C}^{(1)}(\lambda)}\tau)= \sum_{k=0}^n\frac{1}{k+1}\sum_{l=0}^k(-1)^j\phi^{(k-l)}_0(\tau)
\otimes_{{\Bbb C}^{(1)}(\lambda)}\phi^{(l)}_0(\tau),
\quad \tau \in V_0
$$
transferred to $\oplus_{k=0}^n B_{k0}$ gives rise to the coupled BKP hierarchy
\beq
\label{BKPHce}
\sum_{p=0}^k\sum_{j\in {\Bbb N}^{\rm{odd}}}S_j(2y_j)S_j(-\frac2j D_j)
\left(\exp\sum_{s\in {\Bbb N}^{\rm{odd}}}y_sD_s\right)\tau_p\tau_{k-p}\quad k=0,\dots,n.
\eeq
For example the first non trivial ones (which therefore 
can be viewed as   generalisation  to the $B$ case of those written by
Hirota) are the coefficients of $y_6$ in the expansion of \rref{BKPHce}:
$$
\sum_{p=0}^k(D_1^6-5D_1D_3-5D^2_3+D_1D_5)\tau_p\tau_{k-p}=0\quad k=0,\dots,n.\\
$$
Performing the change of variables $w_0=2\frac{\partial\log (\tau_0)}{\partial x_1}$ and 
$w_1=\frac{\tau_i}{\tau_0}$ $ i=1,\dots n$
these equations become
$$
\begin{array}{ll}
(w_{0xxxxx}+30w_{0x}w_{0xxx}-5w_{0xxy}-30w_{0x}w_{0y}+
60w^3_{0x}+9w_{0t})_x-5w_{0yy}=0&\\
-5w_{1yy}+180w_{0x}w_{1xx}+9w_{1xt}+30w_{0xxx}w_{1xx}&\\
+30w_{0x}w_{1xxxx}+w_{1xxxxxx}-30w_{0x}w_{1xy}-30w_{0y}w_{1xx}-5w_{1xxxy}=0&
\end{array}
$$
where $x=x_1$ $y=x_3$ $t=x_5$. Once again from these equations by performing opportune 
reduction process (namely eliminating the variable $x_{(2m+1)j}$) we can obtain the coupled 
$B_m$ soliton equations. In particular for $m=1$ we get the coupled Kotera--Sawada
hierarchy \cite{SKo}, whose first non trivial equation when $n=1$  
$w_0=2(\log(\tau_0))_{xx}$ and 
$w_1=\frac{\tau_1}{\tau_0}$ is
$$
\begin{array}{ll}
&9w_{0t}+w_{0xxxxx}+3w_{0x}w_{0xxx}+3w_{0}w_{0xxx}+180w^2_0w_{0x}=0\\
&180w_{0}w_{1xx}+9w_{1xt}+30w_{0xx}w_{1xx}+30w_{0}w_{1xxxx}
+w_{1xxxxxx}=0
\end{array}
$$
where $x=x_1$ and $t=x_5$.\par\noindent
Of course exactly as in the non coupled case these hierarchies can be also obtained
by applying our construction to the Lie ``polynomial'' algebras $(A^{(2)}_1)^{(n)}$. 
While for $m=2$ one obtains the coupled $B_2$ hierarchies, which again when $n=1$ has as 
first non trivial equation:
$$
\begin{array}{ll}
(w_{0xxxxx}+30w_{0x}w_{0xxx}-5w_{0xxt}-30w_{0x}w_{0t}+60w^3_{0x})_x-5w_{0tt}=0&\\
-5w_{1tt}+180w_{0x}w_{1xx}+30w_{0xxx}w_{1xx}+30w_{0x}w_{1xxxx}&\\
+w_{1xxxxxx}
-30w_{0x}w_{1xt}-30w_{0t}w_{1xx}-5w_{1xxxt}=0&
\end{array}
$$
where $x=x_1$ and $t=x_3$.
Finally the vertex operator construction provides (as in the case of the hierarchies 
of type $A$) the multi--soliton solutions for all the hierarchies written above.\\\\
{\large{\bf Acknowledgements}}\par 
The authors wish thank Franco Magri for many helpful discussions and for 
the interest showed for  our work.
P.C. would like to thanks very warmful the participants of the
Conference on Infinite Dimensional Lie Theory and Its Applications
(July 21-25, 2003) held in The Fields Institute of Toronto
for very interesting discussions which have stimulated 
 the first ideas for the present work and the members of Mathematics Department
of the Chalmers in G\"oteborg where part of the work has  been done for their 
welcomed hospitality.
G.O. thanks Sergio Cacciatori for useful discussions on the physical relevance
of the integrable hierarchies.

\end{document}